\documentclass[journal, 10pt]{IEEEtran}
\IEEEoverridecommandlockouts
\usepackage{tabularx}
\usepackage{mathrsfs} 
\usepackage{multirow}
\usepackage{color, colortbl}

\usepackage[binary-units]{siunitx} 
\DeclareSIUnit{\bps}{bps}
\usepackage{algorithm2e}
\usepackage{amsmath,amssymb,amsfonts}
\usepackage{cite}
\usepackage{graphicx} \usepackage[caption=false,font=footnotesize]{subfig}
\usepackage{bbm}
\usepackage{amsthm}
\usepackage{comment}

\newtheorem{remark}{Remark}

\begin{document}

\title{A Frequency-Domain Opportunistic Approach for Spectral-Efficient Cell-Free Massive MIMO}
\author{Wei~Jiang,~\IEEEmembership{Senior~Member,~IEEE}
        and Hans~Dieter~Schotten,~\IEEEmembership{Member,~IEEE}
\thanks{W. Jiang and H. D. Schotten are with the Intelligent Networking Research Group, German Research Centre for Artificial Intelligence (DFKI), Kaiserslautern, Germany, and are also with University of Kaiserslautern  (RPTU), Paul-Ehrlich Street, Kaiserslautern, 67663 Germany (e-mail: wei.jiang@dfki.de; schotten@eit.uni-kl.de).}
}

\maketitle

\begin{abstract}
Constrained by weak signal strength and significant inter-cell interference, users located at the cell edge in a cellular network suffer from inferior service quality. Recently, cell-free massive MIMO (CFmMIMO) has gained considerable attention due to its capability to offer uniform quality of service, alleviating the cell-edge problem. In contrast to previous studies focused on narrow-band CFmMIMO systems, this paper studies wideband CFmMIMO communications  against channel frequency selectivity. By exploiting the frequency-domain flexibility offered by orthogonal frequency-division multiplexing (OFDM), and leveraging a particular spatial characteristic in the cell-free structure -- namely, the near-far effect among distributed access points (APs) -- we propose an opportunistic approach to boost spectral efficiency. The core concept lies in opportunistically activating nearby APs for certain users across their assigned OFDM subcarriers while deactivating distant APs to prevent power wastage and lower inter-user interference. Furthermore, this approach enables the use of downlink pilots by reducing the number of active APs per subcarrier to a small subset, thereby substantially improving downlink performance through coherent detection at the user receiver.  Verified by numerical results, our proposed approach demonstrates considerable performance improvement compared to the two benchmark approaches.
\end{abstract}

\begin{IEEEkeywords}
Cell-free massive MIMO, coherent detection, downlink pilot, frequency domain, max-min power control, near-far effect, opportunistic communications, OFDM, user-centric 
\end{IEEEkeywords}

\IEEEpeerreviewmaketitle

\section{Introduction}

In a conventional cellular network, a base station (BS) is positioned at the center of each cell. Cell-center users, situated closer to the BS, generally enjoy superior quality of service (QoS), whereas those at the cell edge often face poor QoS due to weak signal strength caused by distance-dependent path loss, strong inter-cell interference, and inherent handover issues in cellular architecture. The performance gap between the cell center and edge is substantial, rather than trivial. For reference, \SI{30}{\bps\per\hertz{}} in the downlink and \SI{15}{\bps\per\hertz{}} in the uplink are the minimum peak spectral efficiency (SE) for the fifth generation (5G) system, as specified in ITU-R M.2410 recommendation \cite{Ref_non2017minimum}.  On the other hand,  the $5^{th}$ percentile SE, alternatively termed the $95\%$-likely rate \cite{Ref_ngo2017cellfree} or $5\%$-outage rate \cite{Ref_nayebi2017precoding}, serves as the performance metric for the cell edge. In contrast to high peak SE, the targets set by ITU-R for the $5^{th}$ percentile SE are significantly lower, standing at \SI{0.3}{\bps\per\hertz{}} (downlink) and \SI{0.21}{\bps\per\hertz{}} (uplink) in indoor hotspots. The requirements are further reduced to \SI{0.12}{\bps\per\hertz{}} and \SI{0.0453}{\bps\per\hertz{}} in rural areas, marking a substantial center-edge performance gap exceeding $100$ times.

Cell-free massive multi-input multi-output (CFmMIMO) \cite{Ref_ngo2017cellfree} has recently attracted a lot of interest from both academia and industry due to its potential to become a key technology for the next generation system \cite{Ref_jiang2021road}. This technology ensures uniform QoS for all users, effectively tackling the under-served issue at the edges of conventional cellular networks  \cite{Ref_jiang2023celledge}. Cells and their boundaries are absent in CFmMIMO. Instead, a large number of distributed low-power, low-cost access points (APs) jointly serve a relatively small number of users over the same time-frequency resource. Like massive MIMO systems with co-located antennas  \cite{Ref_marzetta2015massive}, CFmMIMO is compelled to operate in a time-division duplex (TDD) mode because of excessively high overhead associated with inserting downlink pilots, which scales in proportion to the massive number of service antennas. With TDD, a few users transmit uplink pilots, enabling the network side to estimate the uplink channel response, which is considered equivalent to the downlink channel response due to channel reciprocity. 

Until the present, researchers have accomplished many critical advancements. Ngo et al.  \cite{Ref_ngo2017cellfree} illustrated the effectiveness of CFmMIMO in solving the cell-edge problem and ensuring uniform QoS, through comparing $95\%$-likely rate with that of small cells.
In \cite{Ref_nayebi2017precoding}, Nayebi et al. examined the downlink performance of CFmMIMO across different linear precoding schemes and power optimization algorithms.  Björnson et al. reviewed dynamic cooperation clustering in network MIMO and analyzed the scalability of CFmMIMO \cite{Ref_bjornson2020scalable}. 
The study conducted by Interdonato et al. \cite{Ref_interdonato2019downlink} has shown that employing downlink pilots can improve performance, despite the rise in overhead and additional pilot contamination. In \cite{Ref_ngo2018total}, the energy efficiency of CFmMIMO is investigated, presenting an algorithm to maximize overall energy efficiency while adhering to per-user SE and per-AP power constraints. The authors of \cite{Ref_bjornson2020making} offered a comprehensive analysis of uplink SE across four cell-free implementations, considering different degrees of cooperation among the APs. Buzzi et al. proposed a user-centric (UC) approach \cite{Ref_buzzi2017cellfree, Ref_buzzi2020usercentric},  which can reduce the fronthaul overhead while maintaining comparable performance levels \cite{Ref_ammar2022usercentric}. The synergy of CFmMIMO with other potential 6G technologies has also been explored in the literature, such as reconfigurable intelligent surface (RIS)-aided CFmMIMO discussed in \cite{shi2024ris} and \cite{shi2023uplink}. 

Prior works typically assume flat fading channels. Algorithm design and performance analyses have been carried out within the \textit{coherence interval}, representing the time-frequency duration during which the channel response is approximately considered constant. This assumption holds only true in narrow-band communications. Nevertheless, most wireless systems nowadays are wideband with signal bandwidths far wider than the \textit{coherence bandwidth}, leading to frequency selectivity. The authors of \cite{Ref_jin2020spectral} first examined CFmMIMO across frequency-selective channels. This work uses single-carrier transmission, which lacks support for high data rates due to complex signal equalization \cite{Ref_jiang2016ofdm}. In \cite{Ref_jiang2021cellfree}, we proposed orthogonal frequency-division multiplexing (OFDM)-based multi-carrier transmission for CFmMIMO (CFmMIMO-OFDM) and discussed frequency-domain conjugate beamforming, pilot assignment, and user-specific resource allocation. In \cite{Ref_gao2021uplink}, the uplink transmission design in crowded correlated CFmMIMO-OFDM systems with a limited number of orthogonal pilots is reported.  Later, \cite{Ref_zheng2022cellfree} studied the performance of CFmMIMO-OFDM in high-speed train communications, and revealed that it achieves more uniform QoS than other systems.

To unleash the potential of CFmMIMO-OFDM, this paper proposes a novel technique called opportunistic AP selection (OAS). It aims to boost spectral efficiency by exploiting the frequency-domain degrees of freedom inherent in multi-carrier transmission. Users are spread out across orthogonal frequency resources, with each resource block (RB) accommodating a subset of users. Depending on the number of users per RB, the proposed approach is categorized into two types: single-user OAS (SU-OAS) and multi-user OAS (MU-OAS). Furthermore, this approach exploits a unique spatial characteristic of the cell-free structure -- namely, the near-far effect among distributed APs. The system opportunistically activates a few nearby APs for certain users over their assigned RBs while deactivating distant APs to prevent power wastage and lower inter-user interference. 

The main contributions of this work can be listed as follows:
\begin{itemize}
    \item This article is the first work to utilize the frequency domain to optimize the CFmMIMO systems, introducing a new degree of freedom beyond the conventional narrow-band approaches within a coherence interval. Following the establishment of the CFmMIMO-OFDM signal models, it offers a thorough framework for frequency-domain design, including user assignment, time-frequency pilot pattern, and channel estimation for both downlink and uplink scenarios.
    \item While downlink pilots are unnecessary in co-located massive MIMO due to channel hardening \cite{ngo2017no}, this effect is diminished in CFmMIMO, making downlink pilots useful \cite{interdonato2019downlink}. This article is the first to enable the use of \textit{orthogonal} downlink pilots in CFmMIMO by reducing the number of active APs at each OFDM subcarrier. Furthermore, we demonstrate that coherent detection at the user receiver can significantly enhance downlink performance, rather than trivial.
    \item We derive the closed-form expressions for the achievable SE in both SU-OAS and MU-OAS. To enable a comprehensive comparison, this work extends narrow-band cell-free (CF) and user-centric approaches to wideband OFDM communications and provides their corresponding closed-form SE expressions.
    \item An extensive numerical performance comparison of CF, UC, SU-OAS, and MU-OAS is conducted for both uplink and downlink, considering various factors such as the number of assigned users per OFDM subcarrier, the number of selected APs per user, and different power control strategies, including equal allocation and max-min optimization. 
\end{itemize}

The remainder of this paper is organized as follows: the next section models the CFmMIMO-OFDM system and frequency-selective channels. Sections III and IV introduce the SU-OAS and MU-OAS approaches, respectively. Section V analyzes the performance by deriving their closed-form expressions for SE. Section VI explains the simulation setup and shows some representative numerical results, followed by the conclusions in Section VII. To clarify the novelty of the proposed methods, Table \ref{table_benck} differentiates the four approaches with regard to channel estimation and signal transmission. 
\begin{table}[!tbph]
\renewcommand{\arraystretch}{1.3}
\caption{Comparison of various approaches.}
\label{table_benck}
\centering 
\begin{tabular}{l|c|c|c|c} \hline \hline
 &\multicolumn{2}{c|}{\textbf{Channel Estimation}}& \multicolumn{2}{c}{\textbf{Signal Transmission}}\\ \cline{2-5}
&  Uplink & Downlink & Uplink & Downlink  \\ \hline
CF &$\checkmark$&&all users& all APs \\ \hline
UC &$\checkmark$&& all users& all APs$^\star$ \\ \hline 
\textbf{SU-OAS} &$\checkmark$&$\checkmark$& a single user& a few APs \\ \hline 
\textbf{MU-OAS} &$\checkmark$&$\checkmark$& a few users& a few APs \\ \hline  \hline
\multicolumn{5}{l}{\footnotesize $^\star$Although individual AP only serves its nearby users, collectively,} \\
\multicolumn{5}{l}{\footnotesize all APs involve, as the original CF approach specifies \cite{Ref_buzzi2017cellfree}.}\\
\end{tabular} 
\end{table}

\textit{Notations}: Throughout this article, bold lowercase and uppercase letters represent vectors and matrices, respectively. For their operations,  $(\cdot)^*$, $(\cdot)^T$, $(\cdot)^H$, and $(\cdot)^{-1}$ notate the conjugate, transpose, Hermitian transpose, and inverse of a matrix, respectively, with $[\cdot]_{m,k}$ extracting the $(m,k)^{th}$ entry.  The symbol $\| \cdot\|$ signifies the Frobenius norm, $\lg$ is the decimal logarithm with a base 10 or $\lg=\log_{10}$, while $\circledast$, $\otimes$, and $\odot$ denote the linear convolution, cyclic convolution, and Hadamard product, respectively. $\mathbb{E}$ denotes the statistical expectation, $\mathbf{I}$ stands for an identity matrix of appropriate dimensions, $\lceil x \rceil$ represents the ceiling function, $\bigcup$ and $\bigcap$ are the union and intersection operations of sets, respectively, and the imaginary unit $j$ is defined such that $j^2=-1$.   The tilde $\tilde{}$ marks frequency-domain variables, and the hat $\hat{}$ denotes the estimate of a variable.

\section{System Model}

\begin{figure}[!t]
\centering
\includegraphics[width=0.45\textwidth]{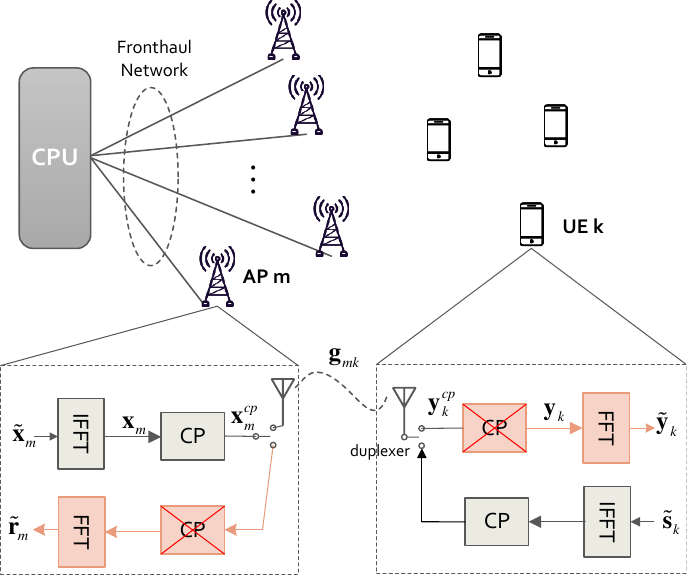}
\caption{System model of a CFmMIMO-OFDM system that consists of a CPU, $M$ APs, and $K$ UEs. It illustrates the block diagram of end-to-end OFDM transmission between AP $m$ and UE $k$, where the DFT demodulator and IDFT modulator are implemented by fast Fourier transform (FFT) and inverse FFT (IFFT), respectively, when $N$ is a power of two. } 
\label{Diagram_SysModel}
\end{figure}

In a cell-free system, unlike cellular networking which divides the network into cells, there are no cells or cell boundaries  \cite{Ref_ngo2017cellfree}. Instead, a large number of $M$ single-antenna APs are randomly distributed to cover a relatively isolated area, such as factories, stadiums, shopping malls, airports, railway stations, exhibition halls, islands, or small towns. A central processing unit (CPU) is connected to all APs through a fronthaul network, coordinating their service to $K$ active users equipped with a single antenna, as depicted in \figurename~\ref{Diagram_SysModel}. We assume perfect fronthaul that offers error-free and sufficient capacity \cite{Ref_bashar2019maxmin}.  The sets of APs and users are denoted by $\mathbb{M}= \{1,\ldots,M\}$ and $\mathbb{K}=\{1,\ldots,K\}$, respectively.  
To harness the benefits of channel hardening and favorable propagation, the number of service antennas must significantly exceed the number of users \cite{Ref_marzetta2010noncooperative}, i.e., $M\gg K$. This allows linear signal processing techniques to perform nearly as well as nonlinear methods, such as the optimal dirty-paper coding \cite{Ref_costa1983writing}. It is worth noting that $M$ refers to the total number of service antennas, rather than the number of APs, particularly when multi-antenna APs\footnote{ 
Signal models, performance analyses, and closed-form SE expressions in this article can be directly applied to a general scenario involving multi-antenna APs, provided that the co-located antennas at each AP experience independent channel fading. This extension is achieved by treating a multi-antenna AP as multiple single-antenna APs that share an identical value of large-scale fading, as demonstrated by a unified CFmMIMO model with a varying number of antennas per AP in \cite{Ref_jiang2024unified}. However, in multi-antenna AP scenarios, it is essential to consider the channel correlation among co-located antennas. Performance analysis under such spatially correlated conditions, as detailed in \cite{Ref_bjornson2020making}, is significantly tedious. For ease of illustration, this article uses single-antenna APs, leaving the study of multi-antenna APs with correlated channels for future work. } 
are employed.

\subsection{Frequency-Selective Channel Model}
Previous CFmMIMO studies were carried out under the assumption that channels undergo frequency-flat fading, modeled by a scalar. For example, \cite{Ref_ngo2018total, Ref_ngo2017cellfree, Ref_bjornson2020scalable, Ref_nayebi2017precoding, Ref_interdonato2019downlink, Ref_masoumi2020performance, Ref_bjornson2020making, Ref_buzzi2017cellfree, Ref_buzzi2020usercentric, Ref_ammar2022usercentric, Ref_zheng2022cellfree, Ref_jiang2021impactcellfree, Ref_zeng2021pilot} utilize a circularly symmetric complex Gaussian random variable with zero mean and unit variance, i.e., $h \sim \mathcal{CN}(0, 1)$, to model small-scale fading. This assumption holds true only within the context of narrow-band communications. However, modern mobile systems, since the advent of third generation systems in the early 2000s, are characterized by wideband communications, which suffer from severe frequency selectivity. From a practical perspective, this article takes a step forward in its focus on \textit{frequency-selective fading} channels in wideband communications. 

The small-scale fading for the frequency-selective channel between AP $m$ and user $k$ at the $t^{th}$ OFDM symbol is modeled as a linear time-varying filter in a baseband equivalent basis  \cite{Ref_tse2005fundamental}, i.e.,  
\begin{equation}
    \mathbf{h}_{mk}[t] = \Bigl[ h_{mk,0}[t],\ldots,h_{mk,l}[t],\ldots, h_{mk,L_{mk}-1}[t]  \Bigr]^T,
\end{equation}
where the filter length $L_{mk}$ should be no less than the multi-path delay spread $T_{d,mk}$ normalized by the sampling interval $T_s$. In other words, it should satisfy $L_{mk} \geqslant \left \lceil \frac{T_{d,mk}}{T_s} \right \rceil  $. The gain for each tap $l=0,\ldots,L_{mk}-1$ is calculated by \cite{Ref_jiang20226GbookChapter10}
\begin{equation}
    h_{mk,l}[t] = \sum_i a_i(tT_s)e^{-2\pi j f_c \tau_i(tT_s)} \mathrm{sinc}\left[l - \frac{\tau_i(tT_s)}{T_s}\right],
\end{equation}
where $f_c$ represents the carrier frequency, $a_i(tT_s)$ and  $\tau_i(tT_s)$ denote the discrete-time attenuation and delay of the $i^{th}$ signal path, respectively, and the sinc function $\mathrm{sinc}(x)\triangleq \frac{\sin(x)}{x}$ for $x\neq 0$. 

We use $\beta_{mk}$ to represent the large-scale channel fading between AP $m$ and user $k$, incorporating path loss and shadowing effects.  By combining small-scale and large-scale fading, the channel linking AP $m$ and user $k$ can be modeled as
\begin{align} \nonumber
\mathbf{g}_{mk}[t]&=\Bigl[ g_{mk,0}[t],\ldots,g_{mk,l}[t],\ldots, g_{mk,L_{mk}-1}[t]  \Bigr] ^T\\
&=\sqrt{\beta_{mk}} \mathbf{h}_{mk}[t],
\end{align}
given $g_{mk,l}[t]=\sqrt{\beta_{mk}} h_{mk,l}[t]$, $\forall l$. 
With respect to the channels, as did prior works like \cite{Ref_ngo2018total, Ref_bjornson2020scalable, Ref_interdonato2019downlink, Ref_masoumi2020performance, Ref_bjornson2020making, Ref_buzzi2017cellfree, Ref_buzzi2020usercentric, Ref_ammar2022usercentric, Ref_zheng2022cellfree, Ref_jiang2021impactcellfree, Ref_zeng2021pilot}, our assumptions are as follows:
\begin{enumerate}
    \item \textit{a priori knowledge of large-scale fading} --- Large-scale fading experiences much slower variations and remains constant at least for tens of OFDM symbol periods, depending on the user's mobility. AP $m$ measures and delivers $\{\beta_{mk} \mid k\in \mathbb{K}\}$, the CPU gets global channel knowledge, i.e., $ \{ \beta_{mk} \mid k\in \mathbb{K},\: m\in \mathbb{M}\}$. As $\beta_{mk}$ is frequency-independent, its acquisition and distribution are not hard to implement. Therefore, we assume that $\beta_{mk}$, $\forall m\in\mathbb{M}$ and $k\in\mathbb{K}$, are perfectly known \textit{a priori} and subsequently used at each AP and UE for precoding, power control, and signal detection. 
    \item \textit{independent Rayleigh small-scale fading} --- Assume that $\mathbf{h}_{mk}[t]$ for different pairs of $m\in\mathbb{M}$ and $k\in\mathbb{K}$ are independent.  The rationale behind this assumption is that the APs and UEs are distributed across a wide area, resulting in a distinct set of scatterers for each AP and UE. Each entry of $\mathbf{h}_{mk}[t]$ is a circularly symmetric complex Gaussian random variable with zero mean and unit variance, following the distribution of $\mathcal{CN}(0, 1)$.
\end{enumerate}

\subsection{CFmMIMO-OFDM Signal Model}

To mitigate the significant pilot overhead, which scales with the number of service antennas, massive MIMO systems commonly adopt TDD to separate downlink and uplink signal transmission. Perfect channel reciprocity, which assumes identical channel responses for both uplink and downlink, is considered. Experimental demonstrations have validated the feasibility of accurately calibrating hardware chains to ensure channel reciprocity in CFmMIMO systems \cite{Ref_cao2023experimental}. In the downlink, all APs transmit data symbols over the same time-frequency resource, while the user equipment (UE) for all users simultaneously transmits in the uplink at another instant. 
\subsubsection{Downlink Transmission} 
Write $\tilde{x}_{m}[t,n]$ to denote the transmitted symbol carried on the $n^{th}$ subcarrier of the $t^{th}$ OFDM symbol at AP $m$. OFDM signal transmission is structured in blocks. Forming a frequency-domain symbol block as $\tilde{\mathbf{x}}_m[t] = \left[ \tilde{x}_{m}[t,0],\ldots,\tilde{x}_{m}[t,n],\ldots, \tilde{x}_{m}[t,N-1]  \right]^T$, its covariance matrix satisfies $\mathbb{E}[\tilde{\mathbf{x}}_m \tilde{\mathbf{x}}_m^H ]=p_d\mathbf{I}_N$, where $p_d$ is the power constraint of AP. As shown in \figurename~\ref{Diagram_SysModel}, the OFDM modulator performs an $N$-point inverse discrete Fourier transform (IDFT) to convert $\tilde{\mathbf{x}}_m[t]$ into a time-domain symbol block $\mathbf{x}_m[t] = \left[ x_{m}[t,0],\ldots, x_{m}[t,n'],\ldots, x_{m}[t,N-1] \right]^T$ in terms of 
\begin{equation} \label{OFDMmodulation}
    x_{m}[t, n']=\frac{1}{N}\sum_{n=0}^{N-1}\tilde{x}_{m}[t,n] e^{\frac{2\pi jn'n}{N}},\:\: \forall n'=0,\ldots,N-1.
\end{equation}
Defining the discrete Fourier transform (DFT) matrix as
\begin{equation}
\label {Eqn_DFTMatrix}
\mathbf{D} =
\left[ \begin{aligned}
         \Omega_N^{00} &  &  \cdots && \Omega_N^{0(N-1)} \\
         \vdots &&  \ddots && \vdots \\
         \Omega_N^{(N-1)0} &  &  \cdots && \Omega_N^{(N-1)(N-1)}
\end{aligned} \right]
\end{equation}
with a primitive $N^{th}$ root of unity  $\Omega_N^{n n'}=e^{-\frac{2\pi jn'n}{N}}$, the OFDM modulation in \eqref{OFDMmodulation} is expressed in matrix form as
\begin{equation} \label{eqn:transmitsignal}
   \mathbf{x}_m[t] =\mathbf{D}^{-1} \tilde{\mathbf{x}}_m[t]=\frac{1}{N}\mathbf{D}^{*}\tilde{\mathbf{x}}_m[t]. 
\end{equation}

In multi-path channels, time dispersion leads to not only inter-symbol interference (ISI) from the preceding symbol but also inter-carrier interference among OFDM subcarriers. A guard interval known as a cyclic prefix (CP) is inserted between two consecutive blocks to eliminate ISI and preserve subcarrier orthogonality \cite{Ref_peled1980frequency}. Repeating the last portion of $\mathbf{x}_m[t]$ and adding it at the beginning, yields 
\begin{align}
    &\mathbf{x}_m^{cp}[t]=\\ \nonumber
    & \biggl[
               \underbrace{x_{m}[t, N{-}L_{cp}], \ldots,  x_{m}[t,N{-}1]}_{CP}, x_{m}[t,0], \ldots, x_{m}[t,N{-}1] \biggr ]^T.
\end{align}
The ISI can be eliminated if the length of CP  is no less than the length of any channel filter, i.e., $L_{cp} \geqslant \max \left( L_{mk}\right)$, for all $m\in\mathbb{M}$ and $k\in\mathbb{K}$.  

The use of CP transforms the \textit{linear convolution}, which models the effect of a signal passing through a wireless channel, into a \textit{circular convolution}, facilitating straightforward frequency-domain processing \cite{Ref_jiang2016ofdm}. Assume the downlink signal transmission across the entire network is \textit{time-synchronized}, as justified by the remark at the end of this part.  Therefore, the received symbol block at user $k$ can  be expressed as  
\begin{equation} \label{eQN_RxSignalwithCP}
    \mathbf{y}_k^{cp}[t]=\sum_{m=1}^M  \mathbf{g}_{mk}[t]\circledast \textbf{x}_m^{cp}[t] +\mathbf{z}_k[t],
\end{equation}
where $\mathbf{z}_k[t]$ denotes a vector of additive white Gaussian noise with zero mean and variance $\sigma_z^2$, i.e., $\mathbf{z}_k\sim \mathcal{CN}(\mathbf{0},\sigma_z^2\mathbf{I})$. 
Removing the CP, we get
\begin{equation} \label{eqn:reveivedsignal}
  \mathbf{y}_k[t]=\sum_{m=1}^M  \mathbf{g}_{mk}^N[t]\otimes \mathbf{x}_m[t] +\mathbf{z}_k[t],  
\end{equation}
where $\mathbf{g}_{mk}^N[t]$ represents an $N$-point channel filter formed by padding zeros at the tail of $\mathbf{g}_{mk}[t]$, i.e., 
\begin{equation}
    \mathbf{g}_{mk}^N[t]{=} \biggl[ g_{mk,0}[t],\ldots, g_{mk,L_{mk}-1}[t], \underbrace{0,\ldots,0}_{Zero-padding}  \biggr] ^T.
\end{equation}
 It applies a reasonable assumption that the number of OFDM subcarriers is far larger than the channel filter length, namely $N\gg L_{mk}$. Conducting the OFDM demodulation through the DFT operation, a frequency-domain received symbol block is obtained as
\begin{equation} \label{eqn:frequencydomainRx}
    \tilde{\mathbf{y}}_k[t] = \mathbf{D}\mathbf{y}_k[t].
\end{equation}

Substituting (\ref{eqn:reveivedsignal}) into (\ref{eqn:frequencydomainRx}), and applying \textit{the convolution theorem} \cite{Ref_jiang2016ofdm}, yields
\begin{IEEEeqnarray}{lll}
\label{Eqn_conditionedx} \nonumber
\tilde{\mathbf{y}}_k[t] &=& \sum_{m=1}^{M} \mathbf{D}\left(\mathbf{g}_{mk}^N[t]\otimes\mathbf{x}_m[t]\right)+\mathbf{D}\mathbf{z}_k[t]\\
         & =& \sum_{m=1}^{M} \tilde{\mathbf{g}}_{mk}[t] \odot \tilde{\mathbf{x}}_m[t] + \tilde{\mathbf{z}}_k[t].
\end{IEEEeqnarray}
In this expression, the frequency-domain channel response is defined as $\tilde{\mathbf{g}}_{mk}[t]=[\tilde{g}_{mk}[t,0],\ldots,\tilde{g}_{mk}[t,n],\ldots,\tilde{g}_{mk}[t, N-1]]^T$, obtained through $\tilde{\mathbf{g}}_{mk}[t]=\mathbf{D}\mathbf{g}_{mk}^N[t]$. Similarly, the frequency-domain noise vector \begin{equation}\tilde{\mathbf{z}}_k[t]=[\tilde{z}_{k}[t,0],\ldots,\tilde{z}_{k}[t,n],\ldots,\tilde{z}_{k}[t,N-1]]^T\end{equation} is derived from
  $\tilde{\mathbf{z}}_k[t]=\mathbf{D}\mathbf{z}_k[t]$. 
Decomposing \eqref{Eqn_conditionedx}, we obtain the downlink per-subcarrier model for CFmMIMO-OFDM as
\begin{equation}
\label{Eqn_OFDMDL}
   \tilde{y}_{k}[t,n]=\sum_{m=1}^M \tilde{g}_{mk}[t, n]\tilde{x}_{m}[t,n]+\tilde{z}_{k}[t,n], \:\:k\in\mathbb{K}, 
\end{equation}
where $\tilde{y}_{k}[t,n]$, $\tilde{g}_{mk}[t,n]$, and $\tilde{z}_{k}[t,n]$ are the $n^{th}$ element of $\tilde{\mathbf{y}}_k[t]$, $\tilde{\mathbf{g}}_{mk}[t]$, and $\tilde{\mathbf{z}}_k[t]$, respectively. Thus, a frequency-selective channel is decomposed to a set of $N$ independent frequency-flat subcarriers. 

\begin{remark}
\textit{ The assumption of a synchronized network is practically reasonable. First, the clocks at the APs can be well-synchronized using wired or over-the-air methods \cite{rogalin2014scalable, Ref_interdonato2019ubiquitous}. Compared to co-located massive MIMO, distributed APs induce a larger multipath delay spread due to the largely varying propagation distances across different AP-user pairs, resulting in a channel filter $\mathbf{g}_{mk}[t]$ in \eqref{eQN_RxSignalwithCP} with more taps. However, this asynchronous reception effect does not necessarily impact performance, since the CP in OFDM signals can absorb the combined effects of delay spread and timing misalignment. For example, LTE features a minimal CP length of \( 4.7 \, \mu\text{s} \), corresponding to an excess distance (relative to the shortest propagation path) of \( 1410 \, \text{meters} \), which further increases to around \( 5 \, \text{km} \) for the extended CP of \( 16.67 \, \mu\text{s} \). The asynchronous effect becomes negligible if the length of $\mathbf{g}_{mk}[t]$ remains shorter than the CP. But the coverage area of a CFmMIMO system is probably beyond \( 5 \, \text{km} \); this leads to a significant performance drop \cite{chowdhury2023resilient, li2021impacts, yan2019asynchronous}. Fortunately, OAS proposed in this article ensures that signal transmission occurs only between a few users (or a single user) and a few nearby APs, which substantially reduces delay spread and timing misalignment. The excess propagation distance between a user and its nearby APs can reliably remain within \( 1410 \, \text{meters} \), if a selection threshold is set. As demonstrated in \cite{chowdhury2023resilient}, the synchronization issue of a CFmMIMO system can be effectively alleviated by selecting the nearest AP, achieving a performance comparable to its time-aligned counterpart.} 
\end{remark}

\subsubsection{Uplink Transmission}
Let us turn to the uplink transmission of CFmMIMO-OFDM by writing
\begin{align} \nonumber
    \tilde{\mathbf{s}}_k[t] &= \left[ \tilde{s}_{k}[t,0],\ldots,\tilde{s}_{k}[t,n],\ldots, \tilde{s}_{k}[t,N-1]  \right]^T\\
    \tilde{\mathbf{r}}_m[t] &= \left[ \tilde{r}_{m}[t,0],\ldots,\tilde{r}_{m}[t,n],\ldots, \tilde{r}_{m}[t,N-1]  \right]^T
\end{align} to denote the frequency-domain symbol block transmitted by user $k$ and the frequency-domain symbol block received at AP $m$, respectively. In the uplink, we also assume the signal transmission is \textit{time-synchronized}. 
Analog to \eqref{Eqn_conditionedx}, the uplink transmission can be straightforwardly obtained as 
\begin{align} 
\tilde{\mathbf{r}}_m[t]   & = \sum_{k=1}^{K} \tilde{\mathbf{g}}_{mk}[t] \odot \tilde{\mathbf{s}}_k[t] + \tilde{\mathbf{z}}_m[t],\:m\in\mathbb{M}
\end{align} with $\tilde{\mathbf{z}}_m[t]=[\tilde{z}_{m}[t,0],\ldots,\tilde{z}_{m}[t,n],\ldots,\tilde{z}_{m}[t,N{-}1]]^T$, where $\tilde{\mathbf{z}}_m\sim \mathcal{CN}(\mathbf{0},\sigma_z^2\mathbf{I})$. Thus, the per-subcarrier uplink signal model for CFmMIMO-OFDM is expressed as
\begin{equation} \label{eqn:ULKTranX}
   \tilde{r}_{m}[t,n]=\sum_{k=1}^K \tilde{g}_{mk}[t,n]\tilde{s}_{k}[t,n]+\tilde{z}_{m}[t,n].  
\end{equation}

\section{Single-User Opportunistic AP Selection}
The key idea of OAS lies in spreading out the users across orthogonal frequency resources, where each RB accommodates different subsets of users. Two strategies emerge based on the number of users assigned to each RB: single-user and multi-user opportunistic AP selection. Opting for a single user per RB simplifies the system design by leveraging frequency orthogonality. It eliminates the requirement for complex signal-processing procedures used for mitigating IUI, like MIMO precoding at the transmitter and multi-user detection at the receiver. This section will introduce the SU-OAS approach by detailing the communication process involved in channel estimation, uplink, and downlink data transmission. 

\subsubsection{Frequency-Domain Single-User Assignment}  
\begin{figure}[!t]
\centering
\includegraphics[width=0.44\textwidth]{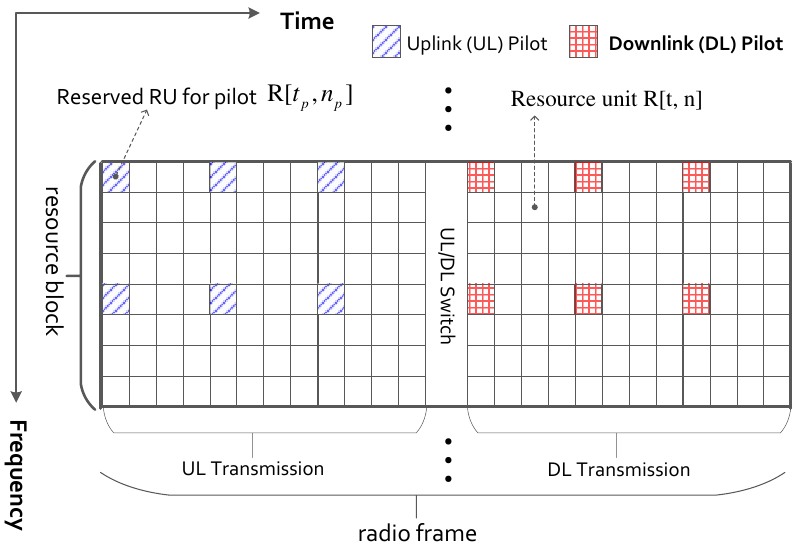}
\caption{The time-frequency resource grid of a CFmMIMO-OFDM system. This figure shows an example where each resource block covers $N_{rb}=8$ subcarriers and spans the entire duration of a radio frame with $T=24$ OFDM symbols. Channel coefficients across all RUs can be acquired through channel estimation and interpolation with the lattice-type pilot pattern. } 
\label{Diagram_OFDMgrid}
\end{figure}

As shown in \figurename~\ref{Diagram_OFDMgrid}, a radio frame is comprised of $T$ OFDM symbols. Write $\mathscr{R} [ t,n ]$ to express a resource unit (RU) offered by the $n^{th}$ subcarrier of the $t^{th}$ OFDM symbol, where $n=0,1,\ldots,N-1$ and $t=0,1,\ldots,T-1$. A resource block is defined as the granularity of resources for allocation, encompassing $N_{rb}$ consecutive subcarriers in the frequency domain and extending throughout the entire duration of a radio frame in the time domain.  There are $B=N/N_{rb}$ (assuming $B$ is an integer) RBs, each of which consists of $T\times N_{rb}$ RUs. The $b^{th}$ RB, labeled by $\mathcal{B}_b$, is mathematically represented as $\mathcal{B}_b \triangleq \left\{ \mathscr{R}[ t,n ] \mid 0\leqslant t \leqslant T-1, \: bN_{rb} \leqslant n<(b+1)N_{rb} \right\}$, where $b$ ranges over $\{0,1,\ldots, B-1\}$. Therefore, the available time-frequency resources for a radio frame can be denoted by $\mathbb{B}=\left\{\mathcal{B}_0,\mathcal{B}_1,\ldots, \mathcal{B}_{B-1} \right\}$. 
User $k$, $\forall k$ periodically reports its QoS request with a scalar $q_{k}$ through the uplink signaling. Then, the CPU knows $\mathbf{q}=\{q_{1},q_{2},\ldots,q_{K}\}$. Utilize $\mathbb{B}_k$ to denote the set of RBs assigned to user $k$. According to some particular criteria, e.g., fairness, priority, and performance, the CPU decides on resource allocation as a function of the users' requests, i.e.,  $\{\mathbb{B}_1,\ldots,\mathbb{B}_K\}=f(\mathbf{q})$, where the implementation of $f(\cdot)$ is out of the scope of this paper and left as a future task. In SU-OAS, a single user is assigned to each RB to simplify the system design by eliminating IUI through frequency orthogonality. The user assignment rule for SU-OAS is mathematically described as $\bigcup_{k=1}^K \mathbb{B}_k \in \mathbb{B}$ (when all RBs are allocated, $\bigcup_{k=1}^K \mathbb{B}_k = \mathbb{B}$), and $\mathbb{B}_k \bigcap \mathbb{B}_{k'}=\varnothing$, $\forall k'\neq k$. Despite the possibility of a user occupying multiple RBs, the perspective from any specific RB maintains the presence of only one user.

\subsubsection{Opportunistic AP Selection} 
\begin{figure}[!t]
\centering
\includegraphics[width=0.45\textwidth]{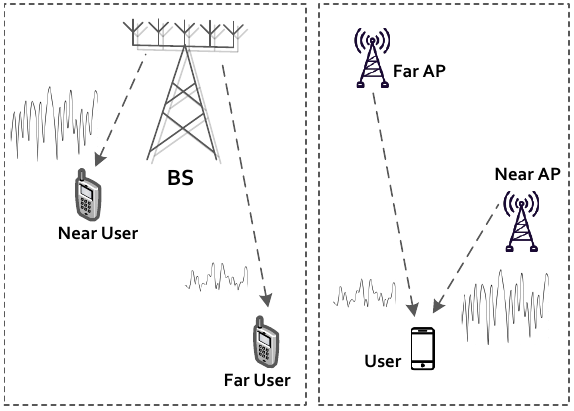}
\caption{Illustration of the near-far effect in a cellular system where a near user gets a strong signal from a BS whereas a far user suffers from a weak signal (left). In comparison, the distributed APs in a cell-free system offer strong and weak signals based on their distances to a user (right). They are labeled as a near AP or a far AP from the perspective of a specific user. } 
\label{Diagram_NFeffect}
\end{figure}

In conventional cellular networks, mobile users located at varying distances from a BS experience different signal strengths. Users in close proximity to the BS are designated as \textit{near users}, while those situated farther away are termed \textit{far users}. It is recognized as a crucial phenomenon called \textit{the near-far effect} in mobile networks.
In cell-free systems, as illustrated in \figurename \ref{Diagram_NFeffect}, the distributed structure gives rise to another form of the near-far effect observed among the distributed APs. From a typical user's standpoint, nearby APs provide strong signal strength, while those from distant APs are weaker. Based on this observation, we categorize the APs into two types: \textit{near APs}, analogous to near users in conventional cellular systems, offering favorable channels with minimal path loss, and \textit{far APs}, where the energy transmitted is inefficiently distributed due to the long propagation distance.

Building on this insight, we introduce the concept of opportunistic AP selection, wherein the transmission of a user's far APs is turned off over its designated RBs. Leveraging the adaptability of the OFDM waveform, turning some RBs off can be seamlessly implemented using virtual subcarriers carrying null symbols \cite{Ref_jiang2016methodUS9319885}. This strategy not only prevents the wasteful energy expenditure by far APs, potentially enhancing both power and spectral efficiency but also alleviates mutual interference by reducing the number of actively transmitting APs. To differentiate near and far APs for each user, several approaches can be conceived, such as
\paragraph{Fixed-Number Selection}
One possible approach is to determine several closest APs for each user in terms of large-scale fading.  
The number of near APs, i.e., $M_s$, is a proper design parameter, where $1\leqslant M_s\leqslant M$. The procedure can be conducted in a centralized manner by the CPU or in a distributed way at each user. For user $k$, sort $ \{ \beta_{mk} \mid m \in \mathbb{M}\}$ in descending order and initially choose the nearest AP in terms of $\mathbbmss{m}_1^k=\arg \max_{m \in \mathbb{M}} \bigl( \beta_{mk}\bigr)$. Then, determine the second nearest AP from the remaining APs as $\mathbbmss{m}_2^k=\arg \max_{m\in \{\mathbb{M}-\mathbbmss{m}_1^k\}} \bigl( \beta_{mk}\bigr)$. This selection process iterates until we obtain the set of near APs as $\mathbb{M}_k=\{\mathbbmss{m}_1^k,\mathbbmss{m}_2^k,\ldots,\mathbbmss{m}_{M_s}^k\}$. 
\paragraph{Threshold-Based Selection}
Another possible approach is to compute a user-specific threshold in terms of the average large-scale fading:
\begin{equation}
    \Bar{\beta}_k=\frac{1}{M} \sum_{m=1}^M \beta_{mk}.
\end{equation}
User $k$ is served by its nearby APs selected in terms of
\begin{equation}
    \mathbb{M}_k = \{m\mid \beta_{mk} \geqslant \epsilon \Bar{\beta}_k \},
\end{equation}
where $\epsilon$ is a coefficient that controls the number of near APs.

\subsubsection{Frequency-Domain Pilots and Channel Estimation} \label{Subsec_pilotandCE}
 
By properly inserting pilot symbols or sequences into the OFDM time-frequency resource grid, channel response can be estimated. The OFDM system generally adopts a lattice pilot design, as illustrated in \figurename~\ref{Diagram_OFDMgrid}. Utilizing both temporal and spectral correlation, the channel coefficient for any RU can be determined by interpolating the channel estimates obtained from the pilots. There are two variants tailored to specific scenarios. In instances of fast fading, comb-type pilots are added at every OFDM symbol to allow the tracking of swift channel changes. Conversely, in situations where channels vary slowly in time but exhibit significant fluctuations in the frequency domain due to strong time dispersion, the block-type pilot arrangement is more effective \cite{Ref_coleri2002channel}.
To focus on key aspects without spending too much effort on known aspects of OFDM channel estimation \cite{Ref_liu2014channel}, this paper employs a \textit{block fading} model. The transmission of a radio frame is carried out within the \textit{coherent time}, and the width of an RB is constrained to be smaller than the \textit{coherence bandwidth}. Hence, the channel coefficients between AP $m$ and user $k$  across all RUs in the $b^{th}$ RB are considered identical, represented by $\tilde{g}_{mk}[b]$. That is    
\begin{equation} \label{eqn:CSIofRB}
    \tilde{g}_{mk}[t,n] = \tilde{g}_{mk}[b],\: \forall \mathscr{R}[t,n] \in \mathcal{B}_b. 
\end{equation}

Conventional CFmMIMO design does not exploit the frequency domain, and therefore relies on time-domain pilots. Due to the limited capacity of the coherence interval, situations arise where multiple users must share the same pilot, resulting in pilot contamination, as discussed in \cite{Ref_zeng2021pilot}. In contrast, CFmMIMO-OFDM distinguishes itself by its capability of enabling more orthogonal pilots through frequency-division multiplexing, leveraging the additional degree of freedom in the frequency domain. Each RB contains $T\times N_{rb}$ RUs, significantly exceeding the number of users. Consequently, it is reasonable to assume the absence of pilot contamination in the CFmMIMO-OFDM system.

\begin{figure}[!bpht]
\centering
\includegraphics[width=0.45\textwidth]{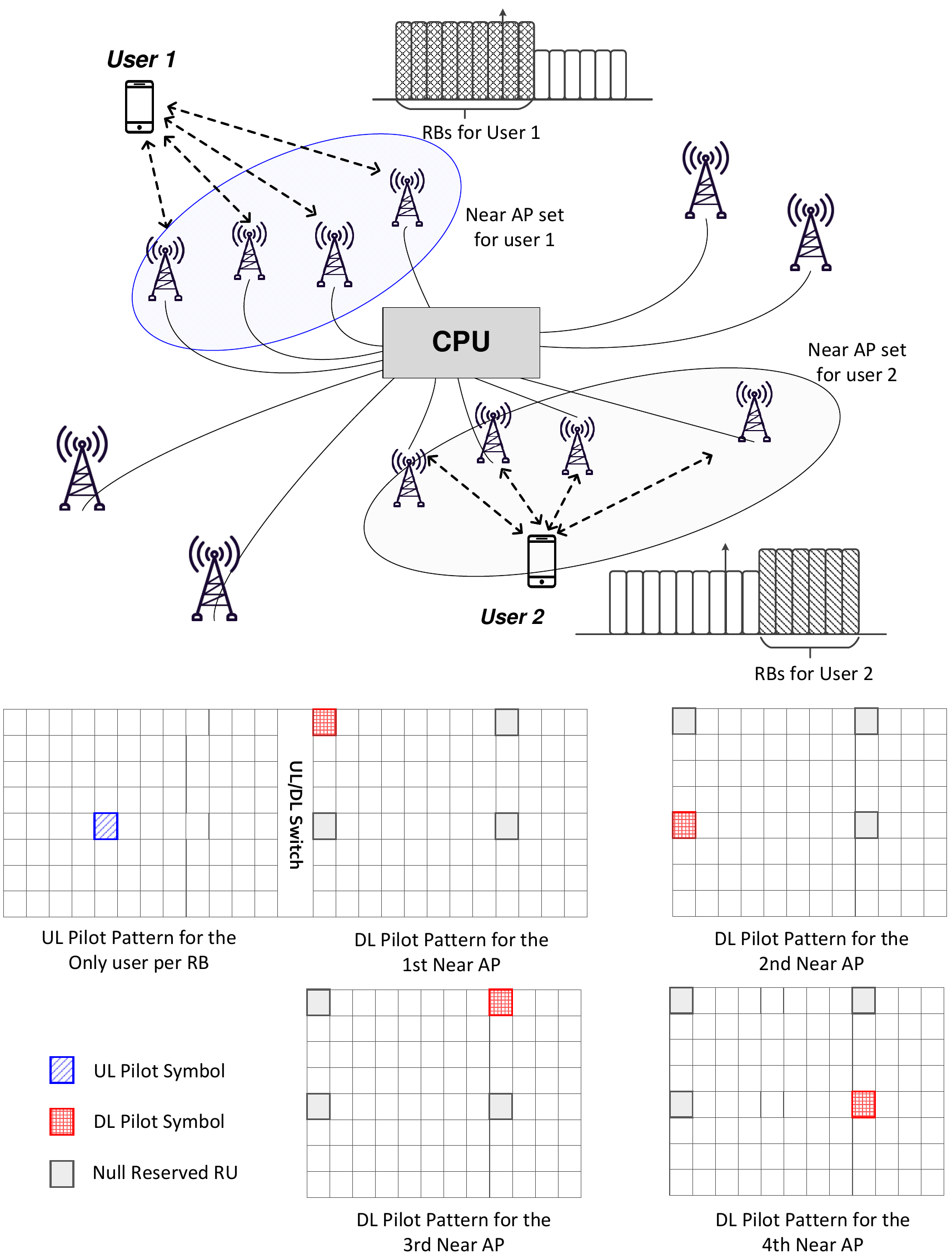}
\caption{Illustration of SU-OAS in a CFmMIMO-OFDM system, along with explanations of the pilot arrangement. To simplify comprehension, we present an example involving two users, each choosing four nearby APs. } 
\label{Diagram_OAS}
\end{figure}

\paragraph{Uplink Channel Estimation} Without loss of generality, let us concentrate on a typical RB to illuminate the uplink channel estimation. When applying the SU-OAS approach, it results in a scenario where communication over $\mathcal{B}_b \in \mathbb{B}_k$ is restricted to user $k$ and its near APs $m\in \mathbb{M}_{k}$. To estimate $\tilde{g}_{mk}[b]$, a pilot symbol is inserted in a reserved RU, denoted by $\mathscr{R} [t_p,n_p] $, in the uplink resource grid of user $k$, as shown in \figurename \ref{Diagram_OAS}. 
The uplink pilot arrangement is expressed by
\begin{equation}
\label{eqn:UL_PilotAssignment} 
\tilde{s}_{k'}[t_p,n_p]=
    \begin{cases}
    \sqrt{p_u}\: \tilde{I},& \text{if } k'=k\\
    0,              & \text{if } k'\neq k
\end{cases}, 
\end{equation}
where $\tilde{I}$ is a known frequency-domain symbol with $\mathbb{E}[\vert \tilde{I} \vert^2]=1$ and $p_u$ denotes the UL transmit power limit. Substituting \eqref{eqn:CSIofRB} and \eqref{eqn:UL_PilotAssignment} into \eqref{eqn:ULKTranX}, yields the received signal of the $m^{th}$ AP on $\mathscr{R} [t_p,n_p] $ as 
\begin{align} \nonumber
   &\tilde{r}_{m}[t_p,n_p] = \sum_{k'=1}^K \tilde{g}_{mk'}[b] \tilde{s}_{k'}[t_p,n_p]+\tilde{z}_{m}[t_p,n_p]\\ \nonumber
   &= \sqrt{p_u}\tilde{g}_{mk}[b] \tilde{I}+ \sqrt{p_u}\sum_{k'\neq k} \tilde{g}_{mk'}[b] \tilde{s}_{k'}[t_p,n_p]+\tilde{z}_{m}[t_p,n_p]\\
                          &=\sqrt{p_u}\tilde{g}_{mk}[b] \tilde{I}+\tilde{z}_{m}[t_p,n_p]. 
\end{align}

Applying the minimum mean-square error (MMSE) estimation obtains an estimate of $\tilde{g}_{mk}[b]$ as
\begin{align} \label{eqn:MMSE} \nonumber
    \hat{g}_{mk}[b] &= \left(\frac{\sqrt{p_u}R_{gg}\tilde{I}^*}{p_uR_{gg}|\tilde{I}|^2 + R_{zz}}\right) \tilde{r}_{m}[t_p,n_p]\\
    &=\left(\frac{\sqrt{p_u}\beta_{mk}\tilde{I}^*}{p_u\beta_{mk}|\tilde{I}|^2 + \sigma_z^2}\right) \tilde{r}_{m}[t_p,n_p],
\end{align}
which applies $R_{gg}=\mathbb{E}\left[ \left \vert \tilde{g}_{mk}[b]\right \vert^2\right]=\beta_{mk}$ and $R_{zz}=\mathbb{E}\left[ \left \vert \tilde{z}_{m}[t_p,n_p]\right \vert^2\right]=\sigma_z^2$.  The variance of the channel estimate is computed as follows:
\begin{align} \nonumber \label{eqn:MMSE_var}
\mathbb{E}\left[\vert\hat{g}_{mk}[b]\vert^2\right]&=\mathbb{E}\left[\hat{g}_{mk}[b]\hat{g}^*_{mk}[b]\right]\\ \nonumber
&=\mathbb{E}\left[\frac{p_u\beta_{mk}^2 \left \vert \tilde{I} \right \vert ^2 \left \vert \sqrt{p_u} \tilde{g}_{mk}[b]\tilde{I}+\tilde{z}_{m}[t_p,n_p]\right \vert ^2}{(p_u\beta_{mk}|\tilde{I}|^2 + \sigma_z^2)^2}\right] \\ \nonumber &=\frac{p_u\beta_{mk}^2\mathbb{E}\left[\left \vert \sqrt{p_u} \tilde{g}_{mk}[b]\tilde{I}+\tilde{z}_{m}[t_p,n_p] \right \vert^2\right]}{(p_u\beta_{mk} + \sigma_z^2)^2}\\
&=\frac{p_u\beta_{mk}^2}{p_u\beta_{mk} + \sigma_z^2}. 
\end{align}
As a result, each near AP $m\in \mathbb{M}_{k}$ gets a channel estimate $\hat{g}_{mk}[b]$. It follows a complex normal distribution $ \mathcal{CN}\left(0,\alpha_{mk}\right)$, where we define $\alpha_{mk}=\frac{p_u\beta_{mk}^2}{p_u\beta_{mk} + \sigma_z^2}$, in comparison with the channel realization $\tilde{g}_{mk}[b]\in \mathcal{CN}(0,\beta_{mk})$. The estimation error raised by additive noise is expressed by $ \xi_{mk}[b] = \tilde{g}_{mk}[b]-\hat{g}_{mk}[b]$, which follows the distribution of $ \mathcal{CN}\left(0, \frac{\sigma_z^2 \beta_{mk}}{p_u\beta_{mk} + \sigma_z^2}\right)$.

\paragraph{Downlink Channel Estimation} A significant challenge in massive MIMO systems, whether cell-free or cellular, is the absence of downlink pilots. That is because the number of orthogonal downlink pilots equals the number of service antennas $M$, which is massive, resulting in prohibitive overhead \cite{Ref_marzetta2015massive}. 
 
Although TDD allows the network to estimate instantaneous CSI $\tilde{g}_{mk}[b]$ during uplink training, users do not have access to this information. Instead, they can only observe the channel statistics $\mathbb{E}[|\tilde{g}_{mk}[b]|^2]$, leading to \textit{non-coherent} detection of the received signals.  In co-located massive MIMO, the law of large numbers ensures that small-scale fading diminishes, meaning the channel becomes hardened, such that
\begin{equation} \label{eQNlimit1}
    \lim_{M \to \infty} \left|\tilde{g}_{mk}[b] \right|^2 = \mathbb{E}[\left|\tilde{g}_{mk}[b] \right|^2].
\end{equation}  
Because the measured channel statistics are sufficiently accurate, downlink pilots are not required \cite{ngo2017no}. 
However, the distributed nature of CFmMIMO results in reduced channel hardening effectiveness. As \eqref{eQNlimit1} is not valid, the difference $|\tilde{g}_{mk}[b] |^2 - \mathbb{E}[|\tilde{g}_{mk}[b] |^2]$ causes channel uncertainty error, thereby justifying the use of downlink pilots. \cite{interdonato2019downlink, kama2024downlink}.    

In particular, for user-centric approaches in \cite{Ref_buzzi2017cellfree} and AP selection-based methods in our work, the number of active APs for each user is quite small, where channel hardening completely vanishes. This further exacerbates the difference between $\left|\tilde{g}_{mk}[b] \right|^2$ and $\mathbb{E}[\left|\tilde{g}_{mk}[b] \right|^2]$, leading to significant performance degradation.
Our proposed scheme involves selectively activating only a few nearby APs for a specific user in its assigned RB. It degrades (virtually) high-dimensional massive MIMO to low-dimensional multi-input single-output (MISO). The number of RUs reserved for downlink pilots equals $M_s$, where $M_s\ll M$, effectively reducing the overhead. It enables inserting downlink pilots and allows users to acquire instantaneous channel estimates, as opposed to relying solely on statistical CSI. Consequently, a fundamental problem limiting the downlink performance of CFmMIMO can be overcome by SU-OAS.

As exemplified by \figurename~\ref{Diagram_OAS}, the downlink pilots for different near APs are orthogonally inserted into reserved RUs,  similar to the mechanism of \textit{cell-specific reference signals} in LTE-Advanced systems for distinguishing multiple antenna ports \cite{Ref_dahlman2011LTE}. A near AP transmits its pilot symbol in a reserved RU, while other APs maintain silence on that particular RU. Denote the reserved RUs as $\mathscr{R} [t_p[m],n_p[m]]  $, where $m\in \mathbb{M}_{k}$.  The first near AP $\mathbbmss{m}_1^k$ sends $\sqrt{p_d}\tilde{I}$ at $\mathscr{R} [t_p[\mathbbmss{m}_1^k],n_p[\mathbbmss{m}_1^k]]  $, but turning off its transmission on other reserved RUs $\mathscr{R} [t_p[m],n_p[m]]  $, $\forall m\neq \mathbbmss{m}_1^k$. The second near AP $\mathbbmss{m}_2^k$ transmits $\sqrt{p_d}\tilde{I}$ on $\mathscr{R} [t_p[\mathbbmss{m}_2^k],n_p[\mathbbmss{m}_2^k]]  $, and keep silence on other reserved RUs $\mathscr{R} [t_p[m],n_p[m]]  $, $\forall m\neq \mathbbmss{m}_2^k$. This rule applies to all other near APs. Consequently, the user's received signal on $\mathscr{R} [t_p[m],n_p[m]] $ is 
\begin{equation}
   \tilde{y}_{k}[t_p[m],n_p[m]]= \sqrt{p_d}\tilde{g}_{mk}[b]\tilde{I}+\tilde{z}_{k}[t_p[m],n_p[m]]. 
\end{equation}
By employing the MMSE estimation as described in \eqref{eqn:MMSE} and \eqref{eqn:MMSE_var}, the user acquires a collection of channel estimates $\{\hat{\mathfrak{g}}_{mk}[b] \mid m\in \mathbb{M}_{k}\}$. These estimates follow a complex normal distribution $ \mathcal{CN}\left(0,\psi_{mk}\right) $, where $\psi_{mk}= \frac{p_d\beta_{mk}^2}{p_d\beta_{mk} + \sigma_z^2}$. Note that $\hat{g}_{mk}$ and $\hat{\mathfrak{g}}_{mk}$ represent two different estimates of $\tilde{g}_{mk}$. The former is accessible at the AP through uplink channel estimation, whereas the latter is obtained by users through downlink channel estimation.

\subsubsection{Uplink Data Transmission} 
Alongside the pilot symbol, user $k$ sends its data payload or control signaling in the uplink. To elaborate, the symbol carried on an uplink RU $\mathscr{R} [t,n]\in \mathbb{B}_k$ is $i_{k}[t,n]$, where $0\leqslant t < T/2$ assuming the ratio between downlink and uplink is $1:1$ for simplicity. This symbol is a unit-variance, independent information-bearing symbol, satisfying $\mathbb{E}[\vert i_{k}[t,n]\vert^2 ]=1$. 
Given the power-control coefficient $0\leqslant \eta_k \leqslant 1$, the transmitted signal from user $k$ can be expressed as follows:
\begin{equation} \label{EQN_SUOAS_UL_SIG}
    \tilde{s}_k[t,n] = \sqrt{p_u \eta_k}  i_k[t,n].
\end{equation}
Since there is only one active user, according to \eqref{eqn:ULKTranX}, the observation of AP $m\in \mathbb{M}_k$ is
\begin{equation} \label{EQN_uplink_Rx_SU}
   \tilde{r}_m[t,n]= \sqrt{p_u  \eta_k} \tilde{g}_{mk}[b] i_k[t,n]+\tilde{z}_{m}[t,n].  
\end{equation}
Each near AP multiplies the received signal with the conjugate of its locally obtained channel estimates and then delivers $\hat{g}_{mk}^*[b]\tilde{r}_m[t,n]$ to the CPU via the fronthaul network. With the pre-processed signals from all near APs, the CPU obtains a soft estimate for $i_k$ as 
\begin{align} 
   \hat{i}_k[t,n] &=\sum_{m\in \mathbb{M}_k} \hat{g}_{mk}^*[b]\tilde{r}_m[t,n] \\ \nonumber
   &= \sum_{m\in \mathbb{M}_k} \hat{g}_{mk}^*[b] \left(\sqrt{p_u  \eta_k} \tilde{g}_{mk}[b] i_k[t,n]+\tilde{z}_{m}[t,n]\right).  
\end{align}

\subsubsection{Downlink Data Transmission} The near APs $m\in \mathbb{M}_k$ jointly transmit the data intended for user $k$ over this user's assigned RBs $\mathbb{B}_k$. The symbol carried on $\mathscr{R} [t,n]\in \mathbb{B}_k$ with $T/2\leqslant t < T$ is $d_{k}[t,n]$, which is a unit-variance, independent information-bearing symbol and therefore satisfies $\mathbb{E}[\vert d_{k}[t,n]\vert^2 ]=1$.  Applying a maximal-ratio method in the frequency domain, the transmitted symbol at AP $m$ is
   \begin{equation} \label{eqn:DLSUOAS_Txsignal}
    \tilde{x}_{m}[t,n] = \sqrt{p_d} \hat{g}_{mk}^*[b] d_{k}[t,n].
\end{equation}
The far APs $m\notin \mathbb{M}_k$ turn off and their transmitted signals can be expressed by
   \begin{equation} \label{eqn:DLMU_TxSyml_forotherAPs}
    \tilde{x}_{m}[t,n] = 0.
\end{equation} 
Thus, the observation of user $k$ can be rewritten from \eqref{Eqn_OFDMDL} as
\begin{equation}
   \tilde{y}_{k}[t,n]= \sqrt{p_d}\sum_{m\in \mathbb{M}_k} \tilde{g}_{mk}[b]  \hat{g}_{mk}^*[b] d_{k}[t,n] +\tilde{z}_{k}[t,n].
\end{equation}

To facilitate a better understanding, the SU-OAS approach is depicted in $\mathrm{Algorithm}$ \ref{alg:CFMIMO001}.
\SetKwComment{Comment}{/* }{ */}
\RestyleAlgo{ruled}
\begin{algorithm} 
\caption{Single-User Opportunistic AP Selection} \label{alg:CFMIMO001}
\ForEach{large-scale fading interval}
{ Measure $\beta_{mk}$, for all $m\in \mathbb{M}$ and $k\in \mathbb{K}$ \;
  Collect QoS request $\mathbf{q}=\{q_{1},q_{2},\ldots,q_{K}\}$ \;
  Assign users $\{\mathbb{B}_1,\ldots,\mathbb{B}_K\}=f(\mathbf{q})$, where  $\bigcup_{k=1}^K \mathbb{B}_k \in \mathbb{B}$ and $\mathbb{B}_k \bigcap \mathbb{B}_{k'}=\varnothing$, $\forall k'\neq k$\;
  \ForEach{User $k\in \mathbb{K}$}{
  Decide the near APs $\mathbb{M}_k=\{\mathbbmss{m}_1^k,\ldots,\mathbbmss{m}_{M_s}^k\}$\;
  }
\ForEach{Resource block $\mathcal{B}_b \in \mathbb{B}_k$, $\forall b$}{
 \For{Uplink slots $t=0:\frac{T}{2}-1$}
 {User $k$ send UL pilot $ \sqrt{p_u}\: \tilde{I}$ at $\mathscr{R} [t_p,n_p] $ and symbol $i_k[t,n]$ at $\mathscr{R} [t,n]$ \;
 Near AP $m\in \mathbb{M}_k$ estimate $\hat{g}_{mk}[b]$ and report $\hat{g}_{mk}^*[b]\tilde{r}_m[t,n]$\;
 CPU detects $\hat{r}[t,n] =\sum_{m\in \mathbb{M}_k} \hat{g}_{mk}^*[b]\tilde{r}_m[t,n]$
 }
  \For{Downlink slots $t=\frac{T}{2}:T-1$}
 { Near AP $m\in \mathbb{M}_k$ send \textbf{DL pilot} $ \sqrt{p_d}\: \tilde{I}$ at $\mathscr{R} [t_p[m],n_p[m]]  $ 
   and precoded symbol $\hat{g}_{mk}^*[b] d_{k}[t,n]$ at $\mathscr{R} [t,n]$ \;
   User $k$ estimate $\hat{\mathfrak{g}}_{mk}[b]$, $\forall m\in \mathbb{M}_{k}$ and detect $\tilde{y}_{k}[t,n]$ \textbf{coherently}\;
 }
  }
   }
\end{algorithm}

\section{Multi-User Opportunistic AP Selection}
The simplicity of SU-OAS comes at the cost of losing the potential to exploit the multiplexing gain, which suggests that the channel capacity for a single-user system is inferior to the sum capacity of a multi-user system \cite{Ref_jiang2023capacity}. Consequently, we propose a multi-user version that assigns more than one user to each RB so as to boost system capacity, albeit at the expense of increased complexity.  This section elaborates on the communication process of MU-OAS in terms of channel estimation, uplink, and downlink data transmission.  

\subsubsection{Frequency-Domain Multi-User Assignment} 
Unlike the single-user assignment, in this scenario, the system assigns multiple users to each RB. We utilize $\mathbb{K}_b=\{\mathbbmss{k}_{b1},\mathbbmss{k}_{b2},\ldots,\mathbbmss{k}_{bN_u}\}\in \mathbb{K}$ to represent the subset of users associated to a typical RB $\mathcal{B}_b$, $\forall b=0,\ldots,B-1$, where $N_u$ is the number of assigned users. Note that a specific user is possible to appear in different RBs based on its QoS request, but this does not affect the system capacity as a whole. This approach increases complexity as it requires precoding at the transmitter and multi-user detection at the receiver to deal with IUI.

\subsubsection{Opportunistic AP Selection}

After the user assignment, the subsequent task is to identify several APs to serve these users over the designated RB. One possible approach involves applying a round-robin scheme where different users $k\in \mathbb{K}_b$ conduct their selection individually.  Similar to the selection strategy in SU-OAS, each user determines the near APs based on large-scale fading. Initially, the first user $\mathbbmss{k}_{b1}$ chooses its nearest AP with the strongest channel in terms of $\mathbbmss{m}_{b1}=\arg \max_{m\in \mathbb{M}} \bigl( \beta_{m{\mathbbmss{k}_{b1}}}\bigr)$. Subsequently, the second user $\mathbbmss{k}_{b2}$ finds its nearest AP as $\mathbbmss{m}_{b2}=\arg \max_{m\in \mathbb{M}} \bigl( \beta_{m{\mathbbmss{k}_{b2}}}\bigr)$. When all users finish the first round, a total of $N_u$ APs are identified. The second round starts when $\mathbbmss{k}_{b1}$ is searching its second AP in terms of $\mathbbmss{m}_{b(N_u+1)}=\arg \max_{m\in \{\mathbb{M}-\mathbbmss{m}_{b1} \} } \bigl( \beta_{m{\mathbbmss{k}_{b1}}}\bigr)$. Likewise, $\mathbbmss{k}_{b2}$ continues to find its second AP as $\mathbbmss{m}_{b(N_u+2)}=\arg \max_{m\in \{\mathbb{M}-\mathbbmss{m}_{b2} \} } \bigl( \beta_{m{\mathbbmss{k}_{b2}}}\bigr)$. This process iterates until the number of selected APs reaches a predefined number of $N_{ap}$. Eventually, we obtain a set of activated APs for $\mathcal{B}_b$ as $\mathbb{M}_b=\{\mathbbmss{m}_{b1},\mathbbmss{m}_{b2},\ldots,\mathbbmss{m}_{bN_{ap}}\}$. It is worth noting that both $N_u$ and $N_{ap}$ are key design parameters. Particularly, it is critical to keep the number of activated APs low. This ensures the practicality of inserting downlink pilots while minimizing unnecessary overhead, which is directly proportional to $N_{ap}$. This distinctive feature sets the proposed scheme apart from the conventional CFmMIMO schemes \cite{Ref_ngo2017cellfree, Ref_buzzi2020usercentric}, where a massive number of APs transmit simultaneously, making downlink pilot-based channel estimation impractical.

\subsubsection{Multi-User Channel Estimation}  
In the uplink, there are now multiple transmitting users. To accommodate orthogonal uplink pilots, the RB $\mathcal{B}_b$ needs to reserve $N_u$ RUs, which are denoted by $\mathscr{R} [t_p[k],n_p[k]]  $, $\forall k\in \mathbb{K}_b$. Similar to the orthogonal pilot arrangement in the downlink of SU-OAS, the first user $\mathbbmss{k}_{b1}$ transmits the pilot symbol $\sqrt{p_u}\tilde{I}$ at $\mathscr{R} [t_p[\mathbbmss{k}_{b1}],n_p[\mathbbmss{k}_{b1}]]  $, while turning off its transmission on other reserved RUs $\mathscr{R} [t_p[k],n_p[k]]  $, $\forall k\neq \mathbbmss{k}_{b1}$. The second user $\mathbbmss{k}_{b2}$ sends $\sqrt{p_u}\tilde{I}$ on $\mathscr{R} [t_p[\mathbbmss{k}_{b2}],n_p[\mathbbmss{k}_{b2}]]  $, while keep silence on other reserved RUs $\mathscr{R} [t_p[k],n_p[k]]  $, $\forall k\neq \mathbbmss{k}_{b2}$. This method applies to all $k\in \mathbb{K}_b$. 
As a result, the received signal of the $m^{th}$ AP on $\mathscr{R} [t_p[k],n_p[k]] $ is 
\begin{equation} 
    \tilde{r}_{m}[t_p[k],n_p[k]] =\sqrt{p_u}\tilde{g}_{mk}[b] \tilde{I}+\tilde{z}_{m}[t_p[k],n_p[k]]. 
\end{equation} 
Applying the MMSE estimation, each AP $m\in \mathbb{M}_{b}$ gets the estimates of its channels with different users $k\in \mathbb{K}_b$, denoted by $\hat{g}_{mk}[b]\in \mathcal{CN}\left(0, \alpha_{mk}\right)$. 
For the downlink of MU-OAS, the pilot arrangement and channel estimation have no difference from that of SU-OAS, as discussed in Sub-Sec. \ref{Subsec_pilotandCE}. 
It is straightforward to derive that each user $k\in \mathbb{K}_b$ obtains the estimates of its channels with $m\in \mathbb{M}_b$, denoted by $\hat{\mathfrak{g}}_{mk}[b]\in \mathcal{CN}\left(0,\psi_{mk}\right)$. 

\subsubsection{Uplink Data Transmission}  
Multiple users simultaneously transmit their signals in the uplink, as opposed to having a single user in SU-OAS. User $k\in \mathbb{K}_b$ intends to deliver the symbol $i_{k}[t,n]$ on the RU for data payload, namely $\mathscr{R} [t,n]\in \mathcal{B}_b$, $\forall t\neq t_p[k]$ and $n\neq n_p[k]$. Since these symbols are unit-variance and independent, the covariance matrix of the vector $\textbf{i} [t,n]=[i_{\mathbbmss{k}_{b1}}[t,n],\ldots, i_{\mathbbmss{k}_{bN_u}}[t,n]]^T$ adheres to $\mathbb{E}[\textbf{i} [t,n]\textbf{i}^H [t,n]]=\mathbf{I}_{N_u}$.  
Unlike the expression in \eqref{EQN_uplink_Rx_SU}, the observation of AP $m\in \mathbb{M}_b$ on $\mathscr{R} [t,n]$ is modified to
\begin{equation} \label{EQnn_mu_ul_RxSigal}
   \tilde{r}_m[t,n]= \sqrt{p_u} \sum_{k\in \mathbb{K}_b} \sqrt{\eta_k}\tilde{g}_{mk}[b] i_k[t,n]+\tilde{z}_{m}[t,n].  
\end{equation}
Each near AP multiplies the received signal with the conjugate of its locally obtained channel estimates and then delivers $\hat{g}_{mk}^*[b]\tilde{r}_m[t,n]$, $\forall k$ to the CPU. To detect the information symbol $i_k$, a soft estimate is formed by combining the user-specific signals from all near APs, i.e.,  
\begin{align} \nonumber \label{QN_ULsoftestimate} 
   \hat{i}_k[t,n] &= \sum_{m\in \mathbb{M}_b} \hat{g}_{mk}^*[b]\tilde{r}_m[t,n] \\ \nonumber
   &= \sqrt{p_u \eta_{k}} \sum_{m\in \mathbb{M}_b} \hat{g}_{mk}^*[b]   \tilde{g}_{mk}[b] i_{k}[t,n] \\ \nonumber
   &+ \sqrt{p_u } \sum_{m\in \mathbb{M}_b} \hat{g}_{mk}^*[b]   \sum_{k'\neq k, k'\in \mathbb{K}_b} \sqrt{\eta_{k'}} \tilde{g}_{mk'}[b] i_{k'}[t,n] \\ 
   &+\sum_{m\in \mathbb{M}_b} \hat{g}_{mk}^*[b]\tilde{z}_{m}[t,n].  
\end{align}

\SetKwComment{Comment}{/* }{ */}
\RestyleAlgo{ruled}
\begin{algorithm} 
\caption{Multi-User Opportunistic AP Selection} \label{alg:CFMIMO002}
\ForEach{large-scale fading interval}
{ Measure $\beta_{mk}$, for all $m\in \mathbb{M}$ and $k\in \mathbb{K}$ \;
  Collect QoS request $\mathbf{q}=\{q_{1},q_{2},\ldots,q_{K}\}$ \;
\ForEach{Resource block $\mathcal{B}_b$, $\forall b$}{
 Assign users $\mathbb{K}_b=\{\mathbbmss{k}_{b1},\ldots,\mathbbmss{k}_{bN_u}\}=f(\mathbf{q})$\;
 Find the near APs $\mathbb{M}_b=\{\mathbbmss{m}_{b1},\ldots,\mathbbmss{m}_{bN_{ap}}\}$\;
 \For{Uplink slots $t=0:\frac{T}{2}-1$}
 {User $k\in \mathbb{K}_b$ send UL pilot $ \sqrt{p_u}\: \tilde{I}$ at $\mathscr{R} [t_p[k],n_p[k]] $ and symbol $i_k[t,n]$ at $\mathscr{R} [t,n]$ \;
 Near AP $m\in \mathbb{M}_b$ estimate $\hat{g}_{mk}[b]$ and report $\hat{g}_{mk}^*[b]\tilde{r}_m[t,n]$\;
 CPU detects $\hat{r}[t,n] =\sum_{m\in \mathbb{M}_b} \hat{g}_{mk}^*[b]\tilde{r}_m[t,n]$
 }
  \For{Downlink slots $t=\frac{T}{2}:T-1$}
 { Near AP $m\in \mathbb{M}_b$ send \textbf{DL pilot} $ \sqrt{p_d}\: \tilde{I}$ at $\mathscr{R} [t_p[m],n_p[m]]  $ 
   and precoded symbol $\sum_{k\in \mathbb{K}_b}\sqrt{\eta_{mk}}\hat{g}_{mk}^*[b] d_{k}[t,n]$ at $\mathscr{R} [t,n]$ \;
   User $k\in \mathbb{K}_b$ estimate $\hat{\mathfrak{g}}_{mk}[b]$, $\forall m\in \mathbb{M}_b$ and detect $\tilde{y}_{k}[t,n]$ \textbf{coherently}\;
 }
  }
   }
\end{algorithm}

\subsubsection{Downlink Data Transmission}  Unlike SU-OAS, where an RU carries one symbol, $\mathscr{R} [t,n]\in \mathcal{B}_b$ in MU-OAS needs to multiplex a group of symbols intended for multiple users, represented by $\mathbf{d}[t,n]=[d_{\mathbbmss{k}_{b1}}[t,n],d_{\mathbbmss{k}_{b2}}[t,n],\ldots,d_{\mathbbmss{k}_{bN_u}}[t,n]]^T$, satisfying $\mathbb{E}[\mathbf{d}[t,n]\mathbf{d}^H[t,n] ]=\mathbf{I}_{N_u}$. 
To spatially multiplex these user-specific symbols, a method called conjugate beamforming (CBF) is generally applied, as \cite{Ref_ngo2018total}.  
The philosophy behind CBF is to amplify the desired signal as much as possible while disregarding interference among users. Extending CBF to the frequency domain, as we proposed in \cite{Ref_jiang2021cellfree}, the transmitted symbol for AP $m\in \mathbb{M}_b$ on $\mathscr{R} [t,n]$ is given by
   \begin{equation} \label{eqn:DLMU_TxSyml}
    \tilde{x}_{m}[t,n] = \sqrt{p_d} \sum_{k\in \mathbb{K}_b} \sqrt{\eta_{mk}} \hat{g}_{mk}^*[b] d_{k}[t,n],
\end{equation} where $0\leqslant \eta_{mk}\leqslant 1$ denotes the power-control coefficient. It indicates how AP $m$ allocates its power among the symbols for different users, which needs to satisfy the per-AP power constraint $\sum_{k\in \mathbb{K}_b} \eta_{mk}\leqslant 1$. Depending on large-scale fading, which is frequency-independent and keeps constant for a relatively long period, $\eta_{mk}$ is independent of $t$ and $n$. 
Substituting \eqref{eqn:DLMU_TxSyml} into \eqref{Eqn_OFDMDL} yields the observation of user $k$ as
\begin{align} \nonumber \label{EQN_dlcbfRxsignal}
   &\tilde{y}_{k}[t,n]=\\
   &\sqrt{p_d}\sum_{m\in \mathbb{M}_b} \tilde{g}_{mk}[b]  \sum_{k'\in \mathbb{K}_b} \sqrt{\eta_{mk'}} \hat{g}_{mk'}^*[b] d_{k'}[t,n] +\tilde{z}_{k}[t,n].
\end{align}
For better comprehension, we summarize the MU-OAS approach in $\mathrm{Algorithm}$ \ref{alg:CFMIMO002}.

\section{Performance Analysis and Comparison}

This section provides a comprehensive performance analysis. We derive the closed-form expressions of achievable SE for the proposed \textit{SU-OAS} and \textit{MU-OAS} approaches, as well as \textit{CF} and \textit{UC}, through adjustments in the numbers of users and APs that participate in signal transmission.

\subsection{Uplink Performance Analysis}
The CPU accesses the full CSI exclusively in two scenarios: first, when channel estimates obtained locally at each AP are transmitted through the fronthaul network, and second, when the CPU performs centralized estimation using received pilot symbols provided by the APs.  Both scenarios incur significant signaling overhead. It is reasonable to assume that the CPU is aware only of channel statistics represented as $ \mathbb{E} [ | \hat{g}_{mk} | ^2]=\alpha_{mk}$, $\forall m, k$, rather than instantaneous channel estimates $\hat{g}_{mk}$, $\forall m, k$. Consequently, the CPU has to detect the received signals in a \textit{non-coherent} manner, aligning with previous works such as \cite{Ref_nayebi2017precoding, Ref_ngo2017cellfree}.  For the sake of simplicity, we omit the time and subcarrier indices of signals, i.e., $[t,n]$ and $[b]$, in the subsequent analysis.

\paragraph{The MU-OAS approach} 

To facilitate the derivation of the signal-to-interference-plus-noise ratio (SINR), \eqref{QN_ULsoftestimate} is decomposed into
\begin{align}  \nonumber \label{XQ_ul_proposed_RxSign}
   \hat{i}_k &= \underbrace{ \sqrt{p_u \eta_k} \sum_{m\in \mathbb{M}_b} \mathbb{E}\left[| \hat{g}_{mk}|^2\right] i_{k}}_{\mathcal{S}_0:\:desired\:signal} + \underbrace{ \sqrt{p_u \eta_k} \sum_{m\in \mathbb{M}_b} \hat{g}_{mk}^*   \xi_{mk} i_{k}}_{\mathcal{I}_1:\:channel-estimation\:error} \\ 
   &+ \underbrace{\sqrt{p_u \eta_k} \sum_{m\in \mathbb{M}_b} (| \hat{g}_{mk}|^2- \mathbb{E}\left[| \hat{g}_{mk}|^2\right] ) i_{k}}_{\mathcal{I}_2:\:channel\:uncertainty\:error} \\  \nonumber 
      &+ \underbrace{\sqrt{p_u } \sum_{m\in \mathbb{M}_b} \hat{g}_{mk}^*   \sum_{k'\neq k, k'\in \mathbb{K}_b} \sqrt{\eta_{k'}}\tilde{g}_{mk'} i_{k'}}_{\mathcal{I}_3:\:inter-user\:interference}  + \underbrace{ \sum_{m\in \mathbb{M}_b} \hat{g}_{mk}^*\tilde{z}_{m}}_{\mathcal{I}_4:\:noise}.  
\end{align}

The terms $\mathcal{S}_0$, $\mathcal{I}_1$, $\mathcal{I}_2$, $\mathcal{I}_3$, and $\mathcal{I}_4$ exhibit mutual uncorrelation. As stated in \cite{Ref_hassibi2003howmuch}, the worst-case noise for mutual information corresponds to Gaussian additive noise with a variance equal to the sum of the variances of $\mathcal{I}_1$, $\mathcal{I}_2$, $\mathcal{I}_3$, and $\mathcal{I}_4$. Thus, the uplink SE for user $k$ is lower bounded by $ \log(1+\gamma_{k}^{mu,ul})$,
 where the effective SINR equals to
\begin{align}  \label{cfmmimo:formularSNR} \nonumber
    \gamma_{k}^{mu,ul}  &= \frac{|\mathcal{S}_0|^2}{\mathbb{E}\left[|\mathcal{I}_1+\mathcal{I}_2+\mathcal{I}_3+\mathcal{I}_4|^2\right]}\\
         &= \frac{|\mathcal{S}_0|^2}{\mathbb{E}\left[|\mathcal{I}_1|^2\right]+\mathbb{E}\left[|\mathcal{I}_2|^2\right]+\mathbb{E}\left[|\mathcal{I}_3|^2\right]+\mathbb{E}\left[|\mathcal{I}_4|^2\right]}.
\end{align} 
The variance of each term can be determined as follows:
\begin{align} 
    |\mathcal{S}_0|^2 & =  p_u\eta_k  \left( \sum_{m\in \mathbb{M}_b} \alpha_{mk}  \right)^2  \\ 
    \mathbb{E}\left[|\mathcal{I}_1|^2\right] & =  p_u\eta_k   \sum_{m\in \mathbb{M}_b} \alpha_{mk} (\beta_{mk}-\alpha_{mk})  \\ 
    \mathbb{E}\left[|\mathcal{I}_2|^2\right] & = p_u \eta_k\sum_{m\in \mathbb{M}_b} \alpha_{mk}^2  \\
    \mathbb{E}\left[|\mathcal{I}_3|^2\right] & =  p_u \sum_{m\in \mathbb{M}_b} \alpha_{mk} \sum_{k'\neq k, k'\in \mathbb{K}_b} \eta_{k'} \beta_{mk'}\\
    \mathbb{E}\left[|\mathcal{I}_4|^2\right] & = \sigma^2_z \sum_{m\in \mathbb{M}_b} \alpha_{mk}. \label{xQN_vairanceresult}
\end{align}
Substituting these variances into \eqref{cfmmimo:formularSNR}, yields  
\begin{equation} \label{GS_SINR_UL_AP}
    \gamma_{k}^{mu,ul} =  \frac{\eta_k \left( \sum_{m\in \mathbb{M}_b} \alpha_{mk}  \right)^2}
    {\sum_{m\in \mathbb{M}_b} \alpha_{mk} \sum_{ k'\in \mathbb{K}_b } \eta_{k'}   \beta_{mk'}  +  \frac{\sigma^2_z}{p_u} \sum_{m\in \mathbb{M}_b} \alpha_{mk}   }.
\end{equation} 

\paragraph{The SU-OAS approach}
It is a specific case of MU-OAS, where the RBs are orthogonally allocated, eliminating IUI and simplifying the system design. By setting $\mathbb{K}_b=\{k\}$ in \eqref{GS_SINR_UL_AP}, the uplink SINR for user $k$ is obtained as
\begin{equation} \label{Gongshi_SuOAS_SE}
    \gamma_{k}^{su,ul} =  \frac{\eta_k \left( \sum_{m\in \mathbb{M}_b} \alpha_{mk}  \right)^2}
    {  \sum_{m\in \mathbb{M}_b}  \eta_{k} \alpha_{mk}     \beta_{mk}  +  \frac{\sigma^2_z}{p_u} \sum_{m\in \mathbb{M}_b} \alpha_{mk}   }.
\end{equation}

Through \eqref{GS_SINR_UL_AP} and \eqref{Gongshi_SuOAS_SE}, we can theoretically compare performance superiority of these two approaches, i.e., 
\newtheorem{theorem}{Theorem}
\begin{theorem}
The SU-OAS approach outperforms the MU-OAS approach in terms of per-user SE.
\end{theorem}
\begin{IEEEproof}
We have 
\begin{align} \nonumber
    \eta_k \beta_{mk}< \eta_k \beta_{mk}+\sum_{k'\neq k, k'\in \mathbb{K}_b } \eta_{k'}   \beta_{mk'} =\sum_{ k'\in \mathbb{K}_b } \eta_{k'}   \beta_{mk'} 
\end{align}
since $\eta_{k}\geqslant 0$ and $\beta_{mk}>0$, $\forall m,k$. Further, 
\begin{equation}
    \sum_{m\in \mathbb{M}_b} \alpha_{mk}\eta_k \beta_{mk}<\sum_{m\in \mathbb{M}_b} \alpha_{mk} \sum_{ k'\in \mathbb{K}_b } \eta_{k'}   \beta_{mk'} 
\end{equation} as $\alpha_{mk}>0$, $\forall m,k$. Then, we get 
\begin{equation} \label{Gongshi_perCom1}
    \gamma_{k}^{su,ul} > \gamma_{k}^{mu, ul}.
\end{equation}
\end{IEEEproof}

\paragraph{The CF approach}
In this case, AP $m$ sees $\tilde{r}_m= \sqrt{p_u } \sum_{k=1}^K \sqrt{\eta_k} \tilde{g}_{mk} i_k+\tilde{z}_{m}$, which consists of signal components from all users, distinguishing it from \eqref{EQnn_mu_ul_RxSigal}.  All $M$ APs take part in the decoding process, where each AP multiplies its received signal with local channel estimates and forwards $\hat{g}_{mk}^*\tilde{r}_m$, $\forall k$ to the CPU. Combining the pre-processed signals from all APs, the CPU gets the soft estimate $ \hat{i}_k = \sum_{m=1}^M \hat{g}_{mk}^*\tilde{r}_m $ for detecting $i_k$. 
When comparing with MU-OAS, we observe that an equivalent analysis process can be applied.  Substituting $\mathbb{M}_b=\mathbb{M}$ and $\mathbb{K}_b=\mathbb{K}$ in the equations from \eqref{XQ_ul_proposed_RxSign} to \eqref{xQN_vairanceresult}, the SINR for user $k$ in CF is obtained as
\begin{equation} \label{GS_SINR_UL_AP_cf}
    \gamma_{k}^{cf, ul} =  \frac{\eta_k \left( \sum_{m=1}^M \alpha_{mk}  \right)^2}
    {\sum_{m=1}^M \alpha_{mk} \sum_{ k'=1}^K \eta_{k'}   \beta_{mk'}  +  \frac{\sigma^2_z}{p_u} \sum_{m=1}^M \alpha_{mk}   }.
\end{equation} 
It is noteworthy that \eqref{GS_SINR_UL_AP_cf} aligns with (27) in \cite{Ref_ngo2017cellfree}, if pilot contamination is absent, which is a reasonable assumption in CFmMIMO-OFDM as discussed in Sub-Sec. \ref{Subsec_pilotandCE}.

\paragraph{The UC approach}

The fundamental concept behind this approach is that each AP only communicates with its proximate users. Specifically, AP $m$ serves a set of top users with the strongest channels, labeled as $\mathcal{K}_m$ \cite{Ref_buzzi2020usercentric}. With knowledge of $\mathcal{K}_m$ for all $m\in\mathbb{M}$, the corresponding set of APs that communicate with user $k$ is determined as $\mathcal{M}_k=\{m \mid k \in \mathcal{K}_m\}$  \cite{Ref_buzzi2017cellfree}.
All users transmit simultaneously, and subsequently, AP $m$ observes $\tilde{r}_m= \sqrt{p_u} \sum_{k=1}^K \sqrt{ \eta_k} \tilde{g}_{mk} i_k+\tilde{z}_{m}$, mirroring the CF approach. However, each AP only pre-processes the receive signal for its associated users. That is, AP $m$, for each $k\in \mathcal{K}_m$, forms  $\hat{g}_{mk}^*\tilde{r}_m$ and forward them to the CPU. 
Upon detecting the symbol from user $k$, the CPU combines the reports from $\mathcal{M}_k$, rather than $\mathbb{M}$ in CF, to form a soft estimate $\hat{i}_k = \sum_{m\in \mathcal{M}_k} \hat{g}_{mk}^*\tilde{r}_m$.
As \eqref{XQ_ul_proposed_RxSign}, we derive the following equation to facilitate SINR calculation:
\begin{align} \nonumber
    \hat{i}_k & = \sqrt{p_u \eta_k} \sum_{m\in \mathcal{M}_k} \mathbb{E}\left[| \hat{g}_{mk}|^2\right] i_{k} + \sqrt{p_u \eta_k} \sum_{m\in \mathcal{M}_k} \hat{g}_{mk}^*  \xi_{mk} i_{k} \\ 
    &+ \sqrt{p_u \eta_k} \sum_{m\in \mathcal{M}_k} (| \hat{g}_{mk}|^2- \mathbb{E}\left[| \hat{g}_{mk}|^2\right] )  i_{k} \\ \nonumber
    &+ \sqrt{p_u } \sum_{m\in \mathcal{M}_k} \hat{g}_{mk}^*  \sum_{k'=1,k'\neq k}^K \sqrt{\eta_{k'}}\tilde{g}_{mk'} i_{k'} + \sum_{m\in \mathcal{M}_k} \hat{g}_{mk}^*\tilde{z}_{m}.
\end{align}

Similar mathematical manipulations as that of \eqref{GS_SINR_UL_AP} result in the subsequent expression for the uplink SINR of user $k$:
\begin{equation}  \label{GS_UCuplink_sinr}
    \gamma_{k}^{uc, ul} =  \frac{\eta_k \left( \sum_{m \in \mathcal{M}_k} \alpha_{mk}  \right)^2}
    {\sum_{m\in \mathcal{M}_k} \alpha_{mk} \sum_{ k'=1}^K \eta_{k'}   \beta_{mk'}  +  \frac{\sigma^2_z}{p_u} \sum_{m\in \mathcal{M}_k} \alpha_{mk}   }.
\end{equation} 
The key difference between UC and MU-OAS lies in the assignment of users. In the MU-OAS approach, only a subset of all users are assigned to each RB. Essentially, the communications occur exclusively between users $k\in \mathbb{K}_b$ and the near APs $m\in \mathbb{M}_b$ over the RB $\mathcal{B}_b$.  By contrast, all users simultaneously associate with every RB in the UC approach since its design does not involve the extra degree of freedom offered by the frequency domain.

Even though \eqref{GS_SINR_UL_AP} and \eqref{GS_UCuplink_sinr} exhibit similarities, their expressions differ. We can establish which one performs better through the following theoretical derivations.  
\begin{theorem}
The MU-OAS approach outperforms the UC approach in terms of per-user SE.
\end{theorem}
\begin{IEEEproof}
To ensure an equitable comparison, identical sets of serving APs for user $k$ are employed in both the UC and MU-OAS approaches, i.e., $\mathcal{M}_k=\mathbb{M}_b$.  Thus, 
\begin{align} \nonumber
    \sum_{ k'\in \mathbb{K}_b } \eta_{k'}   \beta_{mk'} &<  \sum_{ k'\in \mathbb{K}_b } \eta_{k'}   \beta_{mk'}{ +}\sum_{ k'\notin \mathbb{K}_b } \eta_{k'}   \beta_{mk'}\\ & =\sum_{ k'=1}^K \eta_{k'}   \beta_{mk'}
\end{align}
since $\eta_{k}\geqslant 0$ and $\beta_{mk}>0$, $\forall m,k$. Further, 
\begin{equation}
    \sum_{m\in \mathbb{M}_b} \alpha_{mk} \sum_{ k'\in \mathbb{K}_b } \eta_{k'}   \beta_{mk'} < \sum_{m\in \mathcal{M}_k} \alpha_{mk} \sum_{ k'=1}^K \eta_{k'}   \beta_{mk'}
\end{equation} as $\alpha_{mk}>0$, $\forall m,k$. Then, we get 
\begin{equation}
    \gamma_{k}^{mu,ul} > \gamma_{k}^{uc, ul}.
\end{equation}
If combining with \eqref{Gongshi_perCom1}, it is further to know
\begin{equation}
    \gamma_{k}^{su,ul} > \gamma_{k}^{uc, ul}.
\end{equation}
\end{IEEEproof}
This superiority arises from the use of the frequency domain, where only a limited number of users are assigned to each RB, while all users are involved in every RB in the UC (as well as CF) approach.

\subsection{Downlink Performance Analysis}

\paragraph{The MU-OAS approach}

Because of opportunistic selection, only part of APs are activated in a specific RB, while other APs are turned off. This results in the degradation from high-dimensional massive MIMO to low-dimensional MIMO when viewed from the subcarrier's perspective. With users possessing channel estimates enabled by the use of downlink pilots, coherent detection becomes feasible. To detect the information symbol $d_k$, a soft estimate can be reconstructed from \eqref{EQN_dlcbfRxsignal} as follows:
\begin{align} 
\label{Eqn_OFDMdownlink} \nonumber
   \hat{d}_{k}  &=\underbrace{\sqrt{p_d} \sum_{m\in \mathbb{M}_b} \sqrt{\eta_{mk}} \left|\hat{\mathfrak{g}}_{mk}\right|^2 d_{k}}_{  \mathcal{S}_1:}\:desired\:signal\\   
   &+\underbrace{\sqrt{p_d} \sum_{m\in \mathbb{M}_b} \sqrt{\eta_{mk}} \left( \left|\hat{g}_{mk}\right|^2 -\left|\hat{\mathfrak{g}}_{mk}\right|^2 \right)d_{k}}_{\mathcal{I}_5:\:UL-DL\:estimation\:imbalance}\\ \nonumber &+\underbrace{\sqrt{p_d}\sum_{m\in \mathbb{M}_b} \sqrt{\eta_{mk}} \xi_{mk}  \hat{g}_{mk}^* d_{k}}_{\mathcal{I}_1:\:channel-estimation\:error}
    \\ \nonumber &+\underbrace{\sqrt{p_d}\sum_{m\in \mathbb{M}_b} \tilde{g}_{mk} \sum_{k'\neq k, k'\in \mathbb{K}_b}  \sqrt{\eta_{mk'}}  \hat{g}_{mk'}^* d_{k'} }_{\mathcal{I}_3:\:inter-user\:interference}+\underbrace{\tilde{z}_{k}}_{\mathcal{I}_4:\:noise}. 
\end{align}
In contrast to \eqref{XQ_ul_proposed_RxSign}, the channel uncertainty error $\mathcal{I}_2$ arising from the lack of channel estimates vanishes. Nevertheless, there exists an imbalance between the UL and DL estimation for each channel in our approach due to the introduction of downlink pilots. The precoding at the transmitter uses $\hat{g}_{mk}$ whereas the detection at the receiver has to rely on $\hat{\mathfrak{g}}_{mk}$. The new term $\mathcal{I}_5$ reflects this difference.  The variance of each term in \eqref{Eqn_OFDMdownlink} can be determined as follows:
\begin{align} 
    |\mathcal{S}_1|^2 & =  p_d \left(\sum_{m\in \mathbb{M}_b} \sqrt{\eta_{mk} }\psi_{mk}  \right)^2  \\ 
    \mathbb{E}\left[|\mathcal{I}_1|^2\right] & =  p_d   \sum_{m\in \mathbb{M}_b} \eta_{mk} \alpha_{mk} (\beta_{mk}-\alpha_{mk})  \\ 
    \mathbb{E}\left[|\mathcal{I}_3|^2\right] & = p_d \sum_{m\in \mathbb{M}_b} \beta_{mk} \sum_{k'\neq k, k'\in \mathbb{K}_b}  \eta_{mk'}  \alpha_{mk'}\\
    \mathbb{E}\left[|\mathcal{I}_4|^2\right] & =  \sigma^2_z \\
    \mathbb{E}\left[|\mathcal{I}_5|^2\right] & =  p_d   \sum_{m\in \mathbb{M}_b} \eta_{mk} (\psi_{mk}-\alpha_{mk})^2.  
\end{align}
Thus, the SE of user $k$ is expressed by $\log\left(1+\gamma_{k}^{mu,dl} \right)$ with 
\begin{equation} \label{eQQn_SEMUOASDL}
    \gamma_{k}^{mu,dl}=
    \frac{ \left(\sum_{m\in \mathbb{M}_b} \sqrt{\eta_{mk}} \psi_{mk}  \right)^2}
    { 
    \left \{ 
    \begin{aligned}
    &\sum_{m\in \mathbb{M}_b} \beta_{mk} \sum_{ k'\in \mathbb{K}_b}  \eta_{mk'} \alpha_{mk'} \\ 
    &+\sum_{m\in \mathbb{M}_b} \eta_{mk} (\psi_{mk}^2 - 2\psi_{mk}\alpha_{mk})+\frac{\sigma^2_z}{p_d} 
    \end{aligned} \right \} }.
\end{equation}

\begin{remark}    
    To highlight the gain of downlink pilots in our work, non-coherent signal detection without channel estimates is analyzed for comparison. In this case, \eqref{Eqn_OFDMdownlink} is rewritten as
    \begin{align} 
   \nonumber
   \hat{d}_{k}  &=\underbrace{\sqrt{p_d} \sum_{m\in \mathbb{M}_b} \sqrt{\eta_{mk}} \mathbb{E}\left[\left|\hat{\mathfrak{g}}_{mk}\right|^2 \right] d_{k}}_{\mathcal{S}_2:\:desired\:signal}\\   \nonumber
   &=\underbrace{\sqrt{p_d} \sum_{m\in \mathbb{M}_b} \sqrt{\eta_{mk}} \left(\left|\hat{\mathfrak{g}}_{mk}\right|^2 -\mathbb{E}\left[\left|\hat{\mathfrak{g}}_{mk}\right|^2 \right]\right) d_{k}}_{\mathcal{I}_2:\:channel\:uncertainty\:error}\\ 
   &+\mathcal{I}_1+\mathcal{I}_3+\mathcal{I}_4+\mathcal{I}_5.
\end{align}
Since the variance of a sum of independent random variables equals the sum of the variances, we have
\begin{align} \nonumber
\mathbb{E}\left[|\mathcal{I}_2|^2\right] & = p_d \mathbb{E}\left[\left|\sum_{m\in \mathbb{M}_b} \sqrt{\eta_{mk}} \left(\left|\hat{\mathfrak{g}}_{mk}\right|^2 -\mathbb{E}\left[\left|\hat{\mathfrak{g}}_{mk}\right|^2 \right]\right) d_{k}\right|^2 \right] \\ \nonumber
& = p_d \sum_{m\in \mathbb{M}_b} \eta_{mk} \mathbb{E}\left[\left| \left|\hat{\mathfrak{g}}_{mk}\right|^2 -\mathbb{E}\left[\left|\hat{\mathfrak{g}}_{mk}\right|^2 \right] \right|^2 \right] \\ \nonumber
& = p_d \sum_{m\in \mathbb{M}_b} \eta_{mk} \left( \mathbb{E}[ |\hat{\mathfrak{g}}_{mk}|^4 ] - \left (\mathbb{E}[|\hat{\mathfrak{g}}_{mk}|^2 ]  \right)^2 \right) \\
& = p_d \sum_{m\in \mathbb{M}_b} \eta_{mk}   \psi_{mk}^2,
\end{align}
due to the facts that  $\mathbb{E}[ |\hat{\mathfrak{g}}_{mk}|^4 ]=2 \psi_{mk}^2$  (referring to the Appendix) and  $\mathbb{E}[ |\hat{\mathfrak{g}}_{mk}|^2 ]= \psi_{mk}$.
Thus, the SINR of user $k$ in the absence of downlink pilots is expressed by 
\begin{equation} \label{eQQn_SEMUOASDLNODLpilot}
    \Gamma_{k}^{mu,dl}=
    \frac{ \left(\sum_{m\in \mathbb{M}_b} \sqrt{\eta_{mk}} \psi_{mk}  \right)^2}
    { 
    \left \{\begin{aligned}& \sum\limits_{m\in \mathbb{M}_b} \beta_{mk} \sum\limits_{ k'\in \mathbb{K}_b}  \eta_{mk'} \alpha_{mk'} \\ &+\sum_{m\in \mathbb{M}_b}  2 \eta_{mk}  (\psi_{mk}^2 - \psi_{mk}\alpha_{mk})+\frac{\sigma^2_z}{p_d} 
    \end{aligned} \right \}
    }.
\end{equation}
Compared to \eqref{eQQn_SEMUOASDL}, non-coherent detection introduces an additional interference variance of $p_d \sum_{m\in \mathbb{M}_b} \eta_{mk}   \psi_{mk}^2$.
\end{remark}

\paragraph{The SU-OAS approach}
As explained earlier, this approach represents a specific instance of MU-OAS if each RB carries only one user. Consequently, the effective SINR of user $k$ can be obtained by applying $\mathbb{K}_b=\{k\}$ in \eqref{eQQn_SEMUOASDL} as
\begin{equation} 
    \gamma_{k}^{su,dl}=  \frac{ \left(\sum_{m\in \mathbb{M}_b} \sqrt{\eta_{mk}} \psi_{mk}   \right)^2}
    { \Theta_k },
\end{equation}
where $\Theta_k=\sum_{m\in \mathbb{M}_b} \eta_{mk} (\beta_{mk}-\alpha_{mk})   \alpha_{mk}+ \sum_{m\in \mathbb{M}_b} \eta_{mk} (\psi_{mk}-\alpha_{mk})^2+\frac{\sigma^2_z}{p_d}$.

\paragraph{The CF approach}
Here, each AP transmits data symbols intended for all users in the downlink. Therefore, the received signal at user $k$ is given by
\begin{equation}
    \tilde{y}_{k}=   \sqrt{p_d}\sum_{m\in \mathbb{M} } \tilde{g}_{mk}  \sum_{k'\in \mathbb{K}} \sqrt{\eta_{mk'}} \hat{g}_{mk'}^* d_{k'} +\tilde{z}_{k},
\end{equation}
where $\mathbb{K}_b = \mathbb{K}$ and $\mathbb{M}_b = \mathbb{M}$ are substituted into \eqref{EQN_dlcbfRxsignal}. In this traditional scenario, downlink channel estimates cannot be obtained due to the excessive overhead required for orthogonal pilots. As a result, coherent detection becomes infeasible. Instead, a soft estimate based on channel statistics, similar to the approach in \eqref{XQ_ul_proposed_RxSign}, is formed:
\begin{align}
 \nonumber
   \hat{d}_k  &=\underbrace{\sqrt{p_d} \sum_{m\in \mathbb{M}} \sqrt{\eta_{mk}} \mathbb{E} \left [\left|\hat{g}_{mk}\right|^2\right] d_{k}}_{\mathcal{S}_3:\:\text{desired signal}}\\ \nonumber
   &+ \underbrace{\sqrt{p_d} \sum_{m\in \mathbb{M}} \sqrt{\eta_{mk}}\left( \left|\hat{g}_{mk}\right|^2-\mathbb{E} \left [\left|\hat{g}_{mk}\right|^2\right]\right) d_{k}}_{\mathcal{J}_1:\:\text{channel uncertainty error}}  \\ \nonumber &+\underbrace{\sqrt{p_d}\sum_{m\in \mathbb{M}} \hat{g}_{mk} \sum_{k'\in \mathbb{K}, k'\neq k}  \sqrt{\eta_{mk'}}  \hat{g}_{mk'}^* d_{k'} }_{\mathcal{J}_2:\:\text{inter-user interference}}\\ 
   &+\underbrace{\sqrt{p_d}\sum_{m\in \mathbb{M}} \xi_{mk} \sum_{k'\in \mathbb{K} }  \sqrt{\eta_{mk'}}  \hat{g}_{mk'}^* d_{k'}}_{\mathcal{J}_3:\:\text{channel-estimation error}}+\underbrace{\tilde{z}_{k}}_{\mathcal{J}_4:\:\text{noise}}.
\end{align}
Thus, the SINR for user $k$ is given by
\begin{equation} \label{EQN_SINR_of_DL_CF}
    \gamma_{k}^{cf,dl}=  \frac{ \left(\sum_{m\in \mathbb{M}} \sqrt{\eta_{mk}} \alpha_{mk}  \right)^2}
    { \sum_{m\in \mathbb{M}} \beta_{mk} \sum_{k'\in \mathbb{K}}  \eta_{mk'} \alpha_{mk'}+\frac{\sigma^2_z}{p_d} }.
\end{equation}
\begin{IEEEproof}
The derivation of \eqref{EQN_SINR_of_DL_CF} is detailed in the Appendix, which also serves as a reference for deriving other SINR expressions.
\end{IEEEproof}

\paragraph{The UC approach}

The equivalence between UC and CF is established when $\mathbb{M} = \mathcal{M}_k$, as user $k$ is served exclusively by a subset of user-centric APs $\mathcal{M}_k$. As a result, the SINR for user $k$ in the UC approach can be expressed as
\begin{equation} \label{eqn:SNR_SAAP}
    \gamma_{k}^{uc,dl}=  \frac{ \left(\sum_{m\in \mathcal{M}_k} \sqrt{\eta_{mk}} \alpha_{mk}  \right)^2}
    { \sum_{m\in \mathcal{M}_k} \beta_{mk} \sum_{k'\in \mathbb{K}}  \eta_{mk'} \alpha_{mk'}+\frac{\sigma^2_z}{p_d} }.
\end{equation}

\begin{figure*}[!tbph]
\centerline{
\subfloat[]{
\includegraphics[width=0.32\textwidth]{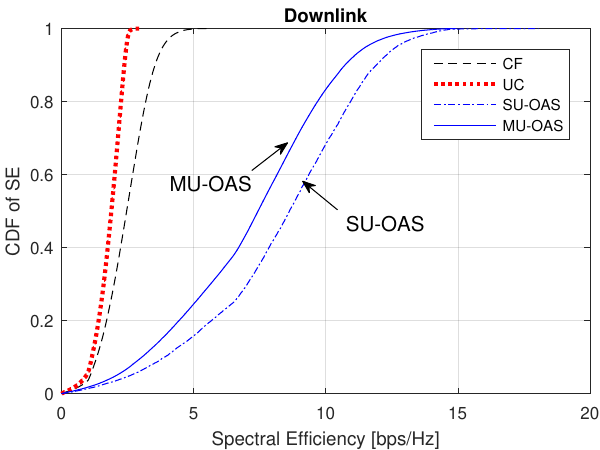}
\label{fig:result1} 
}
\hspace{0mm}
\subfloat[]{
\includegraphics[width=0.32\textwidth]{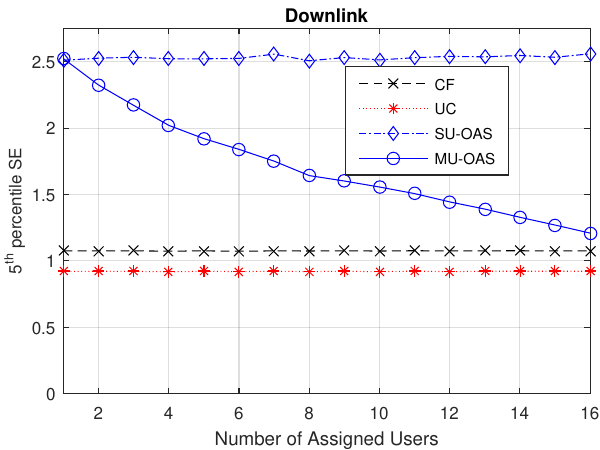}
\label{fig:result2} 
}
\hspace{0mm}
\subfloat[]{
\includegraphics[width=0.32\textwidth]{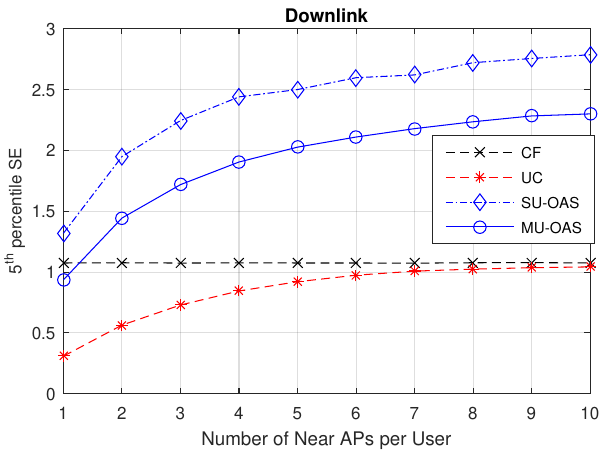}
\label{fig:result3}
} }
\vfill
\centerline{
\subfloat[]{
\includegraphics[width=0.32\textwidth]{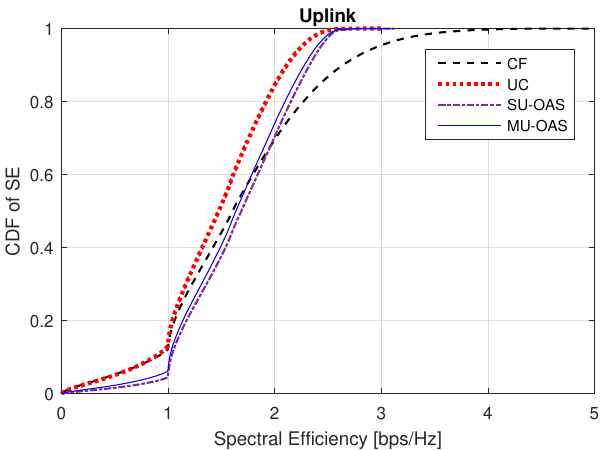}
\label{fig:result4} 
}
\hspace{0mm}
\subfloat[]{
\includegraphics[width=0.32\textwidth]{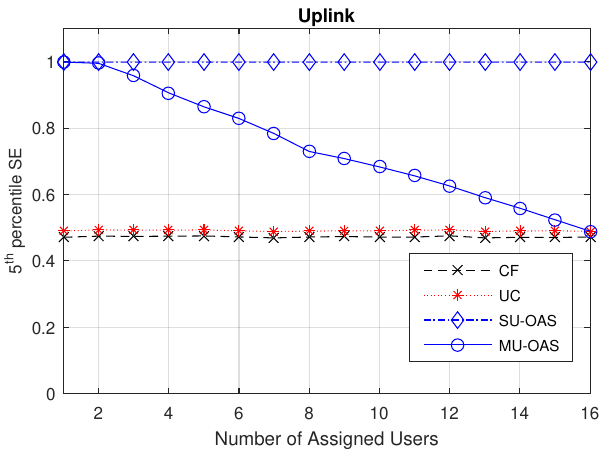}
\label{fig:result5} 
}
\hspace{0mm}
\subfloat[]{
\includegraphics[width=0.32\textwidth]{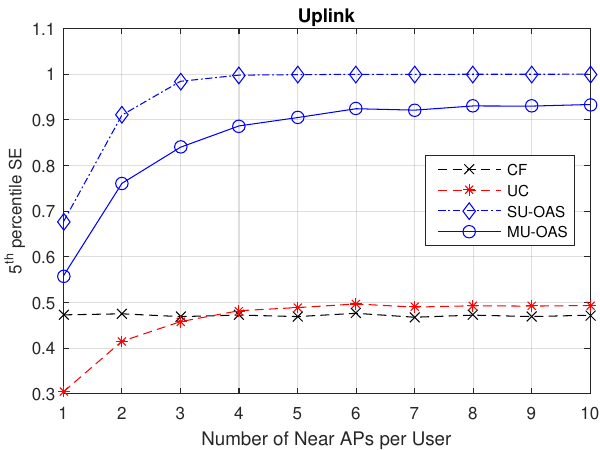}
\label{fig:result6}
}

}
\hspace{0mm}
\caption{Performance comparison of the four approaches in a CFmMIMO-OFDM system, where a total of $256$ antennas are randomly distributed within a circular area with a radius of \SI{1}{\kilo\meter} to serve $16$ users.    }
 \label{diagram_result}
\end{figure*}

\section{Numerical Results}
The gains of the proposed schemes are assessed through numerical evaluation of their achievable spectral efficiency, and comprehensive comparisons with the CF and UC approaches. Unless otherwise specified, our simulations are performed under a default scenario where $M=256$ APs are deployed to serve $K=16$ users. Both the APs and users are randomly distributed within a circular area with a radius of $1\mathrm{km}$. The maximum transmit power for both the APs and UEs are $p_d=0.2\mathrm{W}$ and $p_u=0.2\mathrm{W}$, respectively. The white noise power density equals $-174\mathrm{dBm/Hz}$ with a noise figure of $9\mathrm{dB}$, and the signal bandwidth is  $5\mathrm{MHz}$. It is practical for each UE to utilize a simple full-power strategy, denoted as $\eta_k=1$, $\forall k$, thereby circumventing the necessity for intricate coordination among distributed users.  In the downlink, the APs also employ the full-power strategy. In the case of the CF approach, this implies that $\eta_{m}=\left(\sum_{k=1}^{K} \alpha_{mk} \right)^{-1}$, $\forall m$. For a more thorough comparison, the max-min power optimization \cite{Ref_ngo2017cellfree} is also implemented in CF. In the UC approach, the full-power strategy entails AP $m$ transmitting with $\eta_{m}=\left(\sum_{k\in \mathcal{K}_m} \alpha_{mk} \right)^{-1}$, $\forall m$, as only nearby users are served. Similarly, in the proposed schemes, power is controlled as  $\eta_{m}=\left(\sum_{k\in \mathbb{K}_b} \alpha_{mk} \right)^{-1}$.

Large-scale fading is computed as $\beta_{mk}=10^\frac{\mathcal{L}_{mk}+\mathcal{S}_{mk}}{10}$, where $\mathcal{L}_{mk}$ denotes the path loss in decibel (dB) and the shadowing $\mathcal{S}_{mk}$ follows a log-normal distribution $\mathcal{N}(0,\sigma_{sd}^2)$, with a typical standard deviation of $\sigma_{sd}=8\mathrm{dB}$. Without loss of generality, this article employs the COST-Hata model, as  \cite{Ref_ngo2017cellfree, Ref_ngo2018total, Ref_jiang2021impactcellfree}, which is formulated as  
\begin{equation} \label{eqn:CostHataModel}
    \mathcal{L}_{mk}=
\begin{cases}
-\mathcal{L}_0-35\lg(d_{mk}), &  d_{mk}>d_1 \\
-\mathcal{L}_0-15\lg(d_1)-20\lg(d_{mk}), &  d_0<d_{mk}\leq d_1 \\
-\mathcal{L}_0-15\lg(d_1)-20\lg(d_0), &  d_{mk}\leq d_0
\end{cases},
\end{equation}
where $d_{mk}$ represents the distance between AP $m$ and UE $k$,  $d_0$ and $d_1$ are the breakpoints of the three-slope model and the path loss at the reference distance
\begin{IEEEeqnarray}{ll}
 \mathcal{L}_0=46.3&+33.9\lg\left(f_c\right)-13.82\lg \left(h_{AP}\right)\\ \nonumber
 &-\left[1.1\lg(f_c)-0.7\right]h_{UE}+1.56\lg \left(f_c\right)-0.8
\end{IEEEeqnarray} with the antenna height of AP $h_{AP}$ and the antenna height of UE $h_{UE}$. 
The computation of path loss in \eqref{eqn:CostHataModel} utilizes the following parameters: $d_0=10\mathrm{m}$, $d_1=50\mathrm{m}$, $f_c=2.0\mathrm{GHz}$, $h_{AP}=12\mathrm{m}$, and $h_{UE}=1.7\mathrm{m}$.

First, \figurename \ref{fig:result1} and \figurename \ref{fig:result4} demonstrate the cumulative distribution functions (CDFs) of SE achieved by different schemes in the downlink and uplink, respectively. In MU-OAS, four users (i.e., $N_u=4$) are assigned to each RB by default unless otherwise specified, compared to a single user in SU-OAS. The two proposed approaches, implemented without loss of generality, utilize the fixed-number AP selection with $M_s=5$, where each user selects five nearby APs. For a fair comparison, each user in the UC approach also selects five nearby APs. As anticipated, UC exhibits poorer performance compared with CF in our setting. This is attributed to the fact that only a subset of APs engages in communications. However, this selective engagement brings other advantages, such as reduced power consumption and minimized fronthaul signaling. Their uplink SE exhibits minor differences, as shown in \figurename \ref{fig:result4}, where all uplink schemes rely solely on channel statistics for signal detection. By contrast, the SU-OAS and MU-OAS approaches yield markedly superior downlink SE results compared to the two benchmark approaches. This outcome underscores the effectiveness of incorporating the downlink pilots. Nevertheless, when evaluating the $5^{th}$ percentile SE that signifies uniform service quality, also known as the $95\%$-likely rate in \cite{Ref_ngo2017cellfree} or $5\%$-outage in \cite{Ref_nayebi2017precoding}, the proposed schemes still exhibit significant enhancements in the uplink. The $95\%$-likely rates for CF and UC are approximately \SI{0.49}{\bps\per\hertz^{}}  and \SI{0.50}{\bps\per\hertz^{}}, respectively. In comparison, the $95\%$-likely rate for the SU-OAS approach nearly doubles, reaching \SI{0.99}{\bps\per\hertz^{}}, while the MU-OAS approach achieves nearly \SI{0.92}{\bps\per\hertz^{}}. This superiority of SU-OAS arises from its ability to avoid the IUI through the frequency-domain orthogonality.

To provide further insights, \figurename \ref{fig:result2} illustrates the $5^{th}$ percentile rate of cumulative distribution. Maintaining the same parameters as in the previous simulations, we vary the number of assigned users in MU-OAS from $N_u=1$ to $N_u=16$, aiming to make clear its impact. Since this design parameter solely affects MU-OAS, the curves of the other three approaches remain parallel to the x-axis. As observed, the performance of MU-OAS monotonically decreases with the increasing number of assigned users. When there is only one user, the unique user exclusively enjoys the assigned RBs for the best user-experienced rate, equivalent to SU-OAS. Overall, as highlighted in this figure, the proposed schemes significantly outperform two benchmarks in terms of offering uniform service quality. This trend persists in the uplink, as demonstrated in \figurename \ref{fig:result5}.

Furthermore, \figurename \ref{fig:result3}  and \figurename \ref{fig:result6} present the $5^{th}$ percentile rates concerning different numbers of selected APs per user. Keeping the other parameters consistent, we vary only $M_s$ from $1$ to $10$ in the MU-OAS and SU-OAS approaches. For a fair comparison, each user in the UC approach selects the same number of nearby APs. Given that the CF approach involves all APs, its curves remain parallel to the x-axis. As expected, higher performance is achieved with an increased number of selected APs. However, selecting fewer APs implies reduced power consumption and minimized fronthaul signaling. The optimal choice for striking the best trade-off between performance and energy consumption/fronthaul costs appears to be selecting four or five nearest APs, as evident in these figures. In both downlink and uplink, the proposed scheme's outcomes markedly surpass those of the benchmark approaches. For instance, at $M_s=5$, the $5^{th}$ percentile rates of MU-OAS and SU-OAS in the downlink are elevated to approximately \SI{2}{\bps\per\hertz^{}} and \SI{2.5}{\bps\per\hertz^{}}, respectively, compared to around \SI{1}{\bps\per\hertz^{}} achieved by the two benchmark approaches.

\begin{figure}[!bpht]
\centering
\includegraphics[width=0.4\textwidth]{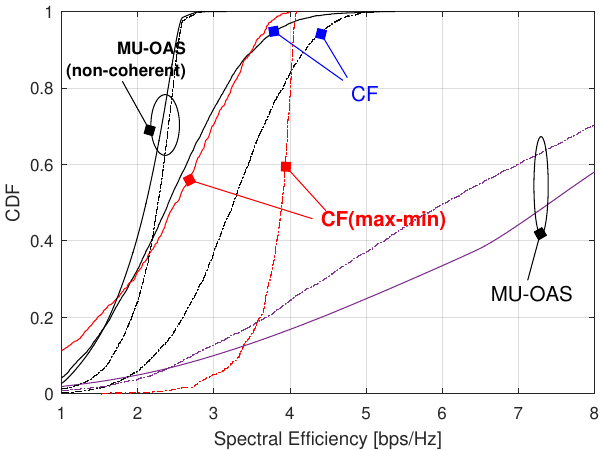}
\caption{  Further comparison of downlink performance with two specific cases: 1) CF using max-min power control, and 2) MU-OAS under non-coherent detection. The solid curves represent the default configuration (1-km radius, $M=256$, and $K=16$), while the dashed-dot curves correspond to an alternative configuration with a 250-meter radius, $M=128$, and $K=8$. }   
\label{Diagram_result_addition} 
\end{figure}

To justify the use of downlink pilots through numerical evaluation, \figurename \ref{Diagram_result_addition} shows the performance of MU-OAS without instantaneous downlink CSI, labeled as \textit{MU-OAS (non-coherent)} in the figure. The noticeable superiority of MU-OAS over its non-coherent counterpart clearly highlights the benefits.
This figure also illustrates the performance of CFmMIMO with max-min power control. However, its impact is not prominent in the default setup. At a specific snapshot $t_1$, this power strategy offers an almost identical rate $R_1$ for all users. When users move to new positions, another uniform rate $R_2$
is achieved. These uniform rates can differ significantly due to the large coverage area of 1000 meters. To address this, we repeated the simulation with a reduced radius of 250 meters, adjusting $M=128$ and $K=8$, respectively. In this configuration, the effect of power control becomes apparent. Although max-min power control can reduce the transmit power of distant APs to zero—akin to deactivating them, as proposed in our method—MU-OAS still achieves superior performance due to the gain of coherent detection. However, using downlink pilots is hard under max-min power control because the activation status of APs is uncertain. Some distant APs may transmit at very low power instead of being fully deactivated, making downlink pilot insertion difficult and inefficient for power amplifiers.
It is worth noting that this power control involves high computational overhead. The average computation time for determining the power coefficients is \num{369.8} seconds on an Intel i7-4790 processor with 3.60GHz and 32GB of memory. This long duration is impractical for scenarios involving user mobility. In contrast, MU-OAS requires only about 0.005 seconds per run, representing a difference of five orders of magnitude in computational efficiency.

Last but not least, it makes sense to emphasize that the performance superiority of the proposed approaches is not necessarily accompanied by the cost of increased complexity. As evident, the process of opportunistic AP selection is simple. Conversely, the decreased number of activated APs and users in the communications can yield additional technical benefits, including reduced power consumption (by turning off far APs), minimized fronthaul overhead, and an alleviated asynchronization effect.

\section{Conclusions}
This paper proposed to opportunistically select the best APs in cell-free massive MIMO systems, leveraging \textit{the near-far} effect among APs, which is a particular feature of the cell-free structure. Exploiting the frequency domain enabled by OFDM, users are first spread out across orthogonal resource blocks. As a result, each resource block carries only a few users, even a single user. Each user selects several nearby APs with strong channels while deactivating the distant APs to save energy and lower interference. From the perspective of a particular resource block, high-dimensional massive MIMO is degraded to low-dimensional MIMO (virtually). Thus, the use of downlink pilots and coherent detection are enabled. Numerical results corroborated that the proposed approaches considerably improve spectral efficiency when compared to the two benchmark approaches, without causing extra complexity.

\appendices

\section{Proof of  \eqref{EQN_SINR_of_DL_CF}}
First, the power gain of the desired signal is given by
\begin{align}  \label{eQn_powerGainofsignal}
    |\mathcal{S}_3|^2 =  p_d \left ( \sum_{m\in \mathbb{M}} \sqrt{\eta_{mk}} \alpha_{mk} \right)^2,
\end{align}
as $\mathbb{E} \left [\left|\hat{g}_{mk}\right|^2\right]=\alpha_{mk}$.
The data symbols and channel realizations are uncorrelated, and $\mathbb{E}[|d_{k}|^2]=1$. Moreover, the variance of a sum of independent random variables equals the sum of their variances.
Thus, the variance of $\mathcal{J}_1$ is determined as follows:
\begin{align}  \label{eQN_deriv0001}
     \mathbb{E}\left[|\mathcal{J}_1|^2\right] & =  p_d \sum_{m\in \mathbb{M}} \eta_{mk} \mathbb{E}\left[\left| \left|\hat{g}_{mk}\right|^2-\mathbb{E}  [|\hat{g}_{mk}|^2] \right|^2 \right] \\ \nonumber
     & =  p_d \sum_{m\in \mathbb{M}} \eta_{mk} \mathbb{E}\left[ |\hat{g}_{mk}|^4- 2\alpha_{mk}|\hat{g}_{mk}|^2  + \alpha_{mk}^2 \right]  \\ \nonumber
     \\ \nonumber
     & =  p_d \sum_{m\in \mathbb{M}} \eta_{mk} \left( \mathbb{E}\left[ |\hat{g}_{mk}|^4\right]- \alpha_{mk}^2 \right).
\end{align}
To proceed, we need to determine $\mathbb{E}\left[ |\hat{g}_{mk}|^4\right]$. Given that $\hat{g}_{mk}$ is a complex Gaussian random variable, specifically $\hat{g}_{mk}\sim \mathcal{CN}(0,\alpha_{mk})$, the power gain $|\hat{g}_{mk}|^2$ follows an exponential distribution with mean $\alpha_{mk}$. Its probability density function is \[f_{|\hat{g}_{mk}|^2}(z) = \frac{1}{\alpha_{mk}} e^{- \frac{z}{\alpha_{mk}}}, \quad z \geq 0. \]
The fourth moment is calculated as
\begin{align} \nonumber
    \mathbb{E}[|\hat{g}_{mk}|^4] &= \mathbb{E}[(|\hat{g}_{mk}|^2)^2] = \int_0^\infty z^2 f_{|\hat{g}_{mk}|^2}(z) \, dz\\ \nonumber
    &=\frac{1}{\alpha_{mk}} \int_0^\infty z^2 e^{-z/\alpha_{mk}} \, dz\\
    &=2 \alpha_{mk}^2.
\end{align}
Substituting this into \eqref{eQN_deriv0001}, we get
\begin{equation}
   \mathbb{E}\left[|\mathcal{J}_1|^2\right] =  p_d \sum_{m\in \mathbb{M}} \eta_{mk} \alpha_{mk}^2. 
\end{equation}
Similarly, the variance of $\mathcal{J}_2$ equals
\begin{align}  \label{eQN_varianceJ2}  
     \mathbb{E}\left[|\mathcal{J}_2|^2\right] & =  p_d \sum_{m\in \mathbb{M}}  \mathbb{E}\left[\left| \hat{g}_{mk} \sum_{k'\in \mathbb{K}, k'\neq k}  \sqrt{\eta_{mk'}}  \hat{g}_{mk'}^*\right|^2 \right] \\ \nonumber
     & =  p_d \sum_{m\in \mathbb{M}}  \mathbb{E}\left[\left| \hat{g}_{mk}\right|^2 \right] \sum_{k'\in \mathbb{K}, k'\neq k}  \eta_{mk'}  \mathbb{E}\left[\left|\hat{g}_{mk'}^*\right|^2 \right]
     \\ \nonumber
     & =  p_d \sum_{m\in \mathbb{M}}  \alpha_{mk} \sum_{k'\in \mathbb{K}, k'\neq k}  \eta_{mk'}  \alpha_{mk'}.
\end{align}
The variance of $\mathcal{J}_3$ is
\begin{align}  \label{eQN_varianceJ03}  
     \mathbb{E}\left[|\mathcal{J}_3|^2\right] & =  p_d \sum_{m\in \mathbb{M}}  \mathbb{E}\left[\left| \xi_{mk} \sum_{ k'\in \mathbb{K} }  \sqrt{\eta_{mk'}}  \hat{g}_{mk'}^*\right|^2 \right] \\ \nonumber
     & =  p_d \sum_{m\in \mathbb{M}}  \mathbb{E}\left[\left| \xi_{mk}\right|^2 \right] \sum_{k'\in \mathbb{K} }  \eta_{mk'}  \mathbb{E}\left[\left|\hat{g}_{mk'}^*\right|^2 \right]
     \\ \nonumber
     & =  p_d \sum_{m\in \mathbb{M}} (\beta_{mk} - \alpha_{mk}) \sum_{k'\in \mathbb{K} }  \eta_{mk'}  \alpha_{mk'}.
\end{align}
Combining these, the total interference power $\mathbb{E}[|\mathcal{J}|^2]=\mathbb{E}[|\mathcal{J}_1|^2] +\mathbb{E}[|\mathcal{J}_2|^2] +\mathbb{E}[|\mathcal{J}_3|^2]$ equals
\begin{align} \nonumber \label{eqn_intpower}
     \mathbb{E}[|\mathcal{J}|^2] &=  p_d \sum_{m\in \mathbb{M}} \eta_{mk} \alpha_{mk}^2 + p_d \sum_{m\in \mathbb{M}}  \alpha_{mk} \sum_{k'\in \mathbb{K}, k'\neq k}  \eta_{mk'}  \alpha_{mk'} \\ \nonumber & + p_d \sum_{m\in \mathbb{M}} (\beta_{mk} - \alpha_{mk}) \sum_{k'\in \mathbb{K} }  \eta_{mk'}  \alpha_{mk'}
   \\ & = p_d \sum_{m\in \mathbb{M}} \beta_{mk} \sum_{k'\in \mathbb{K} }  \eta_{mk'}  \alpha_{mk'}.
\end{align}
With the noise power $\mathbb{E}[|\mathcal{J}_4|^2]  = \sigma^2_z$, and applying \eqref{eQn_powerGainofsignal} and \eqref{eqn_intpower}, we derive \eqref{EQN_SINR_of_DL_CF}.

\bibliographystyle{IEEEtran}
\bibliography{IEEEabrv,Ref_TWC2024}

\begin{IEEEbiography}
[{\includegraphics[width=1in,height=1.25in,clip,keepaspectratio]{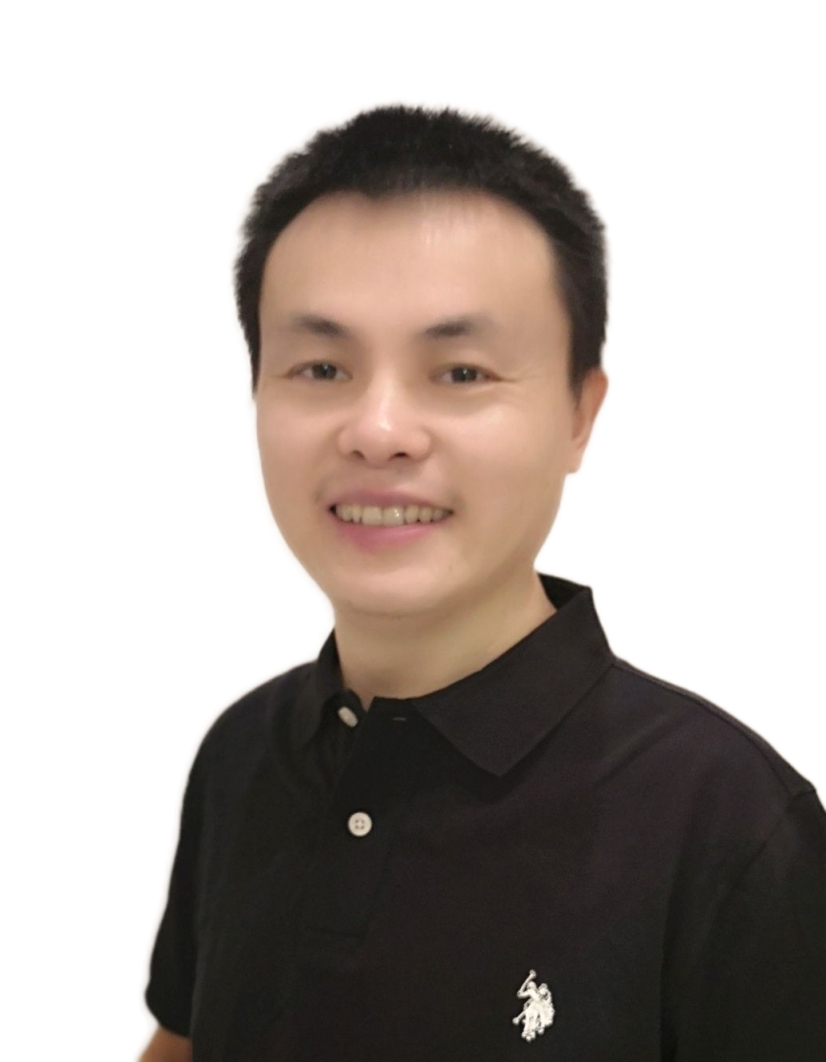}}]
{Wei Jiang} (M'09-SM'19) received a Ph.D. degree in Computer Science from Beijing University of Posts and Telecommunications (BUPT) in 2008. From 2008 to 2012, he was with the 2012 Laboratory, HUAWEI Technologies. From 2012 to 2015, he was with the Institute of Digital Signal Processing, University of Duisburg-Essen, Germany. Since 2015, he has been a Senior Researcher with German Research Center for Artificial Intelligence (DFKI), the biggest European AI research institution and the birthplace of the ``Industry 4.0" strategy. Meanwhile, he was a Senior Lecturer with the University of Kaiserslautern (RPTU), Germany, from 2016 to 2018. He published two monographs: \textit{6G Key Technologies: A Comprehensive Guide (Wiley \& IEEE Press, 2023)} and \textit{Cellular Communication Networks and Standards: The Evolution from 1G to 6G (Springer, 2024)}. He has over 100 papers, holds around 30 granted patents, and led a number of EU and German research projects. He serves as an editor for \textit{IEEE Access} (2019-2023), \textit{IEEE Communications Letters}, \textit{IEEE Open Journal of the Communications Society,} and as a moderator for \textit{IEEE TechRxiv}. He served as a member of the organizing committee or technical committee for many conferences such as ICC, Globecom, ICASSP, VTC, WCNC, and PIMRC. He was the founder and vice chair of the special interest group (SIG) “Cognitive Radio in 5G” under IEEE Technical Committee on Cognitive Networks. 
\end{IEEEbiography}

\begin{IEEEbiography}
[{\includegraphics[width=1in,height=1.25in,clip,keepaspectratio]{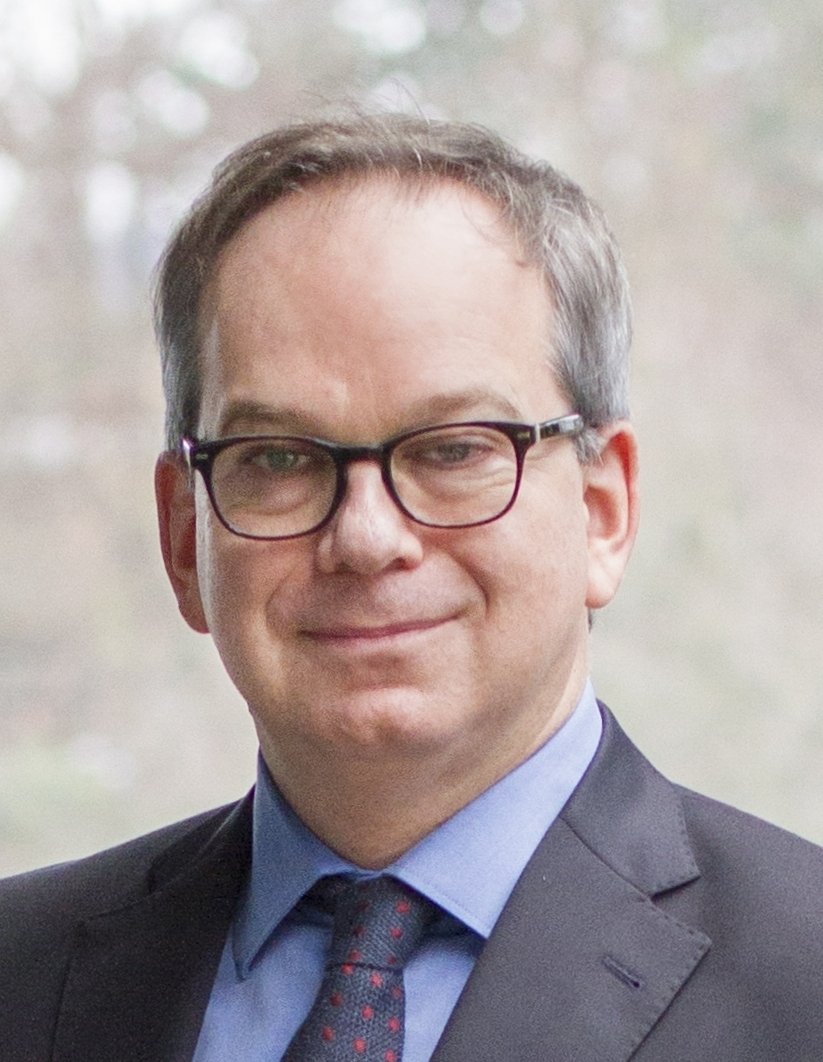}}]
{Hans D. Schotten} (S'93-M'97) received the Ph.D. degree from the RWTH Aachen University of Technology, Germany, in 1997. From 1999 to 2003, he worked for Ericsson. From 2003 to 2007, he worked for Qualcomm. He became manager of a R\&D group, Research Coordinator for Qualcomm Europe, and Director for Technical Standards. In 2007, he accepted the offer to become a full professor at the University of Kaiserslautern. In 2012, he - in addition - became the scientific director of the German Research Center for Artificial Intelligence (DFKI) and head of the Department for Intelligent Networks. Professor Schotten served as dean of the Department of Electrical Engineering of the University of Kaiserslautern from 2013 until 2017. Since 2018, he is chairman of the German Society for Information Technology and a member of the Supervisory Board of the VDE. He is the author of more than 300 papers and participated in 50+ European and national collaborative research projects.
\end{IEEEbiography}
\end{document}